\documentstyle[]{mn}

\input{psfig}

\newcommand{\R}[1]{{\mathrm{#1}}}

\date{Received ???, 1997; accepted ??, ??}

\title[Laser Guide Star for 3.6m and 8m telescopes]{Laser Guide Star for 3.6m and 8m telescopes:\\ Performances and astrophysical implications}
\author[M. Le Louarn et al.]{M.~Le~Louarn$^{1,}$ $^{2}$, R.~Foy$^2$, N.~Hubin$^1$ and M.~Tallon$^2$
\\
$^1$ESO - European Southern Observatory, Karl--Schwarzschild--Stra\ss e 2, Garching bei M\"unchen, D-85748 Federal Republic of Germany\\
$^2$CRAL - Centre de Recherche astronomique de Lyon, 9 av. Charles Andr\'e, F-69561 Saint Genis Laval, France\\
E-mail: lelouarn@eso.org, nhubin@eso.org, foy@obs.univ-lyon1.fr, mtallon@obs.univ-lyon1.fr}

\pagerange{\pageref{firstpage}--\pageref{lastpage}}
\pubyear{1997}

\begin{document}
\maketitle

\label{firstpage}

\begin{abstract}
We  have constructed  an  analytical  model to simulate   the behavior   of an
adaptive optics system coupled  with a sodium laser guide  star.  The code  is
applied to a 3.6-m and 8m class telescopes. The  results are given in terms of
Strehl ratio and full width at half maximum  of the point spread function. Two
atmospheric models are used, one  representing good atmospheric conditions (20
per cent of the time), the other median conditions.

Sky coverage is computed for natural guide star and  laser guide star systems,
with two  different methods.  The first  one is  a statistical approach, using
stellar densities, to compute the probability to find  a nearby reference. The
second  is a  cross-correlation of  a science  object  catalogue and the  USNO
catalogue.  Results are  given in terms of percentage  of the sky that  can be
accessed  with  given performances, and in  terms  of number of science object
that can be observed, with Strehls greater than 0.2 and 0.1 in K and J bands.
\end{abstract}

\begin{keywords}
atmospheric effects -- telescopes
\end{keywords}

\section{Introduction}

The angular resolution of ground based telescopes has been  limited for a long
time  by the atmospheric turbulence.  The  resolution was  typically that of a
telescope 10 -  20 cm in  aperture. The  concept  of adaptive  optics (AO) was
first  proposed by Babcock \shortcite{Babcock1953}.   This technique allows to
correct, in    real time, the   atmospheric  turbulence.  It   provides nearly
diffraction limited images in the near infrared domain.

For the AO system to function,  a reference source must  be found close to the
astronomical object.  It must provide sufficient signal  to noise ratio to the
wave-front   sensor  (WFS) so    that     the wavefront   can  be     measured
accurately. Several other solutions can be used to provide this reference. The
first one is to use the  science object itself, provided  it is bright enough.
The second solution is to use a nearby star.  It has to  be bright enough, but
it must also be  close to  the  science object, within the  isoplanatic patch.
The two previous solutions are called Natural Guide Star  (NGS) AO systems.  A
third solution, proposed  by Foy \&  Labeyrie \shortcite{Foy1985} is  to use a
laser beam to create an artificial laser guide star (LGS).

Two types of laser  stars can be  used, following two principles: Rayleigh and
Mie   scattering  in  the    first  10 to   20  km   of  the  atmosphere  (see
e.g. ~\cite{Fugate1994},~\cite{ONERA})  or resonant scattering by sodium atoms
near 90 km (see e.g. ~\cite{Maxetal1994}).

There are two  main problems with the LGS.   First,the image motion created by
atmospheric turbulence (tip-tilt) can  not be sensed with  the LGS, due to the
round trip of   light. Rigaut \&  Gendron  \shortcite{Rigaut1992}, showed that
this  is a severe  limitation  of the LGS, preventing  full  sky coverage, and
proposed  the concept  of dual  adaptive optics  to increase  the  coverage by
correcting the tilt-reference star itself with a second AO system, to increase
the signal to noise ratio of the  tilt measurement.  Other solutions have been
proposed: one can take   short exposure images,   to freeze the image  motion.
These short   exposure images  are diffraction limited   (because  of the  LGS
correction) and  speckle-type algorithms can then  be used.  The gain compared
to  conventional speckle imaging   is  a gain in  S/N   ratio, due  to the  AO
correction (see Tessier \shortcite{Tessier1997}   for an analysis of  the gain
brought by  short exposure imaging in AO).   The concept  of the polychromatic
artificial star  ~\cite{Foyetal1995},  in   which  the  tilt  information   is
retrieved from laser spots  produced at different wavelengths, gives promising
results ~\cite{Friedman1996} and could solve this problem. Another alternative
would be a combination of different methods relying on  the observation of the
laser spot against the sky background (Ragazzoni \shortcite{Ragazzoni1997}).

The  second  limitation of  the LGS   is the cone  effect   (also called focus
anisoplanatism).  Since the  laser  star is at   a  finite distance  from  the
telescope, compared to the science  object,  the light  coming from the  laser
star  does  not probe  exactly the same  volume of  turbulence  than the light
coming from  the  science object.  This   leads  to a mis-measurement   of the
wavefront, and loss of correction quality.  The cone  effect is reduced if the
artificial  star is located high in  the atmosphere.   Therefore, sodium laser
guide  stars   seem  to   be  better suited   for   astronomical applications.
Approaches  using multiple  guide stars  have  been  proposed  to counter this
effect which is severe on the new 8 to 10 meter class telescopes (e.g.  Tallon
\&  Foy  \shortcite{Tallon1990}, Jankevics \& Wirth \shortcite{Jankevics1991},
Johnston \& Welsh \shortcite{Johnston1994}).

With NGS-AO systems, the sky coverage is limited  to the areas where reference
stars can be  found.  Typically, the sky  coverage is only  a few percents, in
the most optimistic cases. Because of  the tilt determination problem, the sky
coverage of LGS systems must also  be computed.  The goal of  this study is to
compute the sky coverage for two  different systems equipped with sodium laser
guide stars.  The   first is   a   3.6m telescope, representing the    current
generation of instruments, for which a laser guide star upgrade is considered.
The second  system is an 8m  class telescope, representing the new generation,
for which we want to know if a LGS can significantly improve the performances.
The 3.6m class telescope study is representative of an upgraded version of the
ADONIS  (ADaptive Optics   Near  Infra-red  System)  system ~\cite{Beuzit1994}
located in La Silla (Chile).  The 8m class case can be seen as  a model of the
VLT (Very Large Telescope), which is under construction in Cerro Paranal, also
in Chile.  This telescope will  be equipped with  the NAOS (Nasmyth  Adapative
Optics System) system, working in the near  infrared.  The results can be used
however for other systems, since the  atmospheric models can be representative
of a good astronomical site.  In the next section, we describe the models used
to simulate the AO system and the  laser star coupled  with it.  In section 3,
we  estimate  the performances of  these  two  systems in terms  of achievable
Strehl  ratio and Full  Width  at Half Maximum  (FWHM)   of the PSF.  The  sky
coverage is computed,  with  two different  approaches.   The first  one  uses
stellar densities found with synthetic models of our Galaxy.  The second is to
make cross  correlations   of catalogues   containing science   targets  and a
catalogue  containing reference stars.  This  ``observer's approach'' can give
lower limits on the  number of  objects that can  really  be observed, and  in
which  conditions.   In section  4, we   study the  performances  that  can be
expected of an 8m telescope in the red part of the spectrum.  In section 5, we
present our conclusions.

\section{Laser guide star / AO simulation}
\subsection{Atmospheric model}

\begin{figure*}
\centerline{\psfig{figure=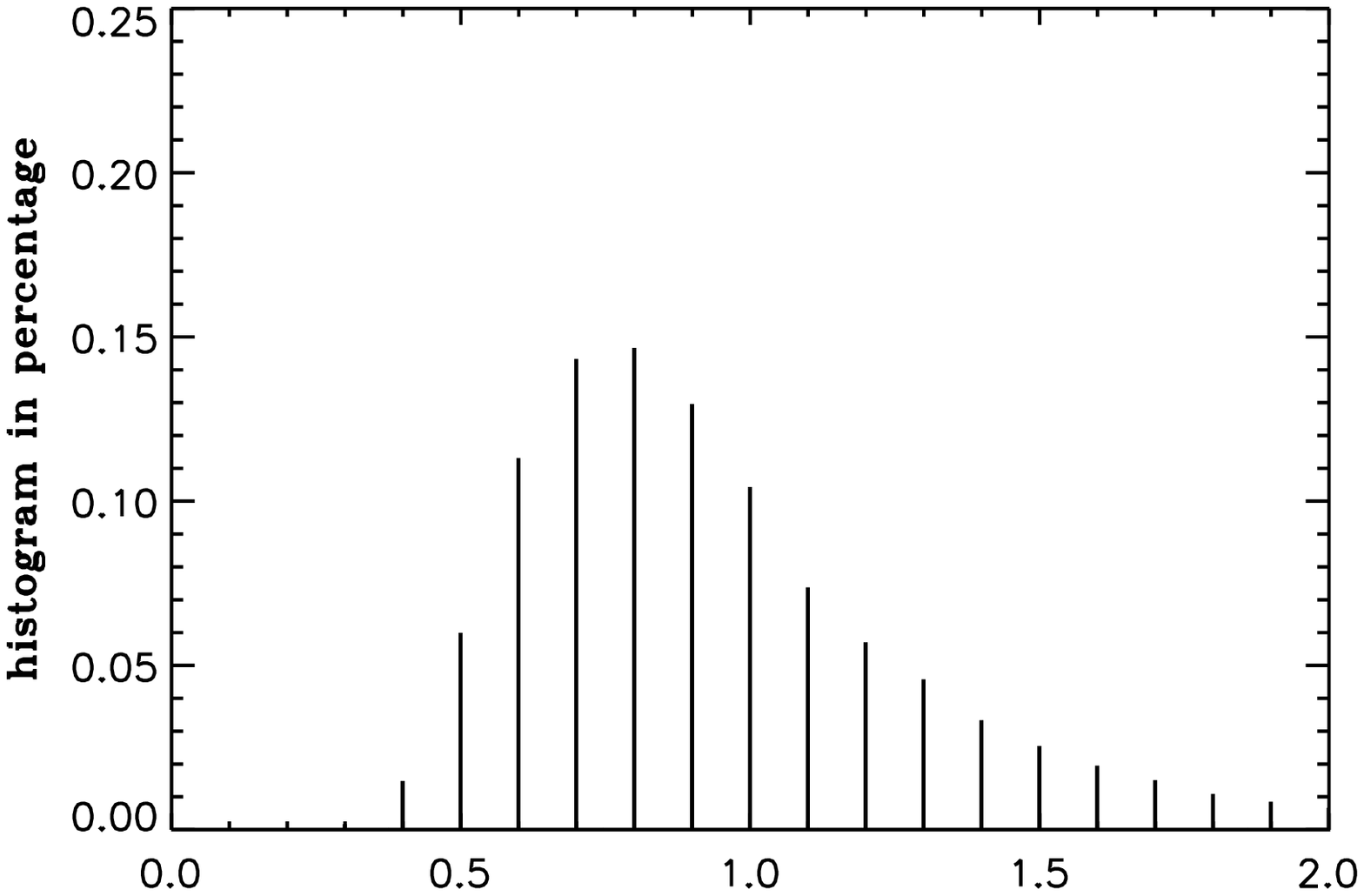,width=9cm}\psfig{figure=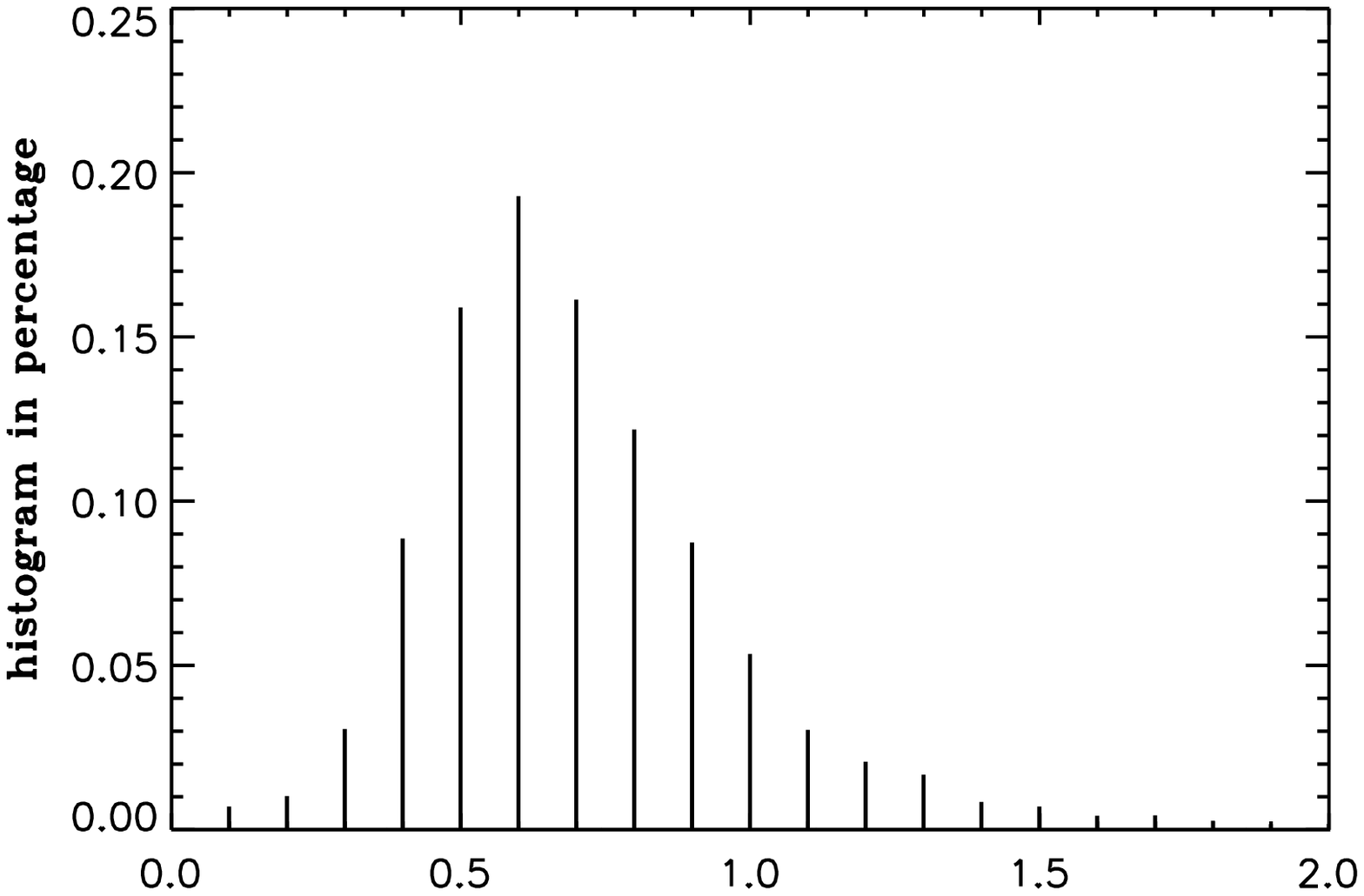,width=9cm}}
\caption{Histogram of seeing (in arcsec) at La Silla (left), Paranal (right)}
\label{fig:seeing_histo}
\end{figure*}

For a simulation of an AO system parameters describing the atmosphere  must be
defined.  The seeing describes the effect that  turbulence  has  on  the long
exposure image.  It is expressed in arc-seconds and represents the FWHM  of the
image, as viewed through an un-compensated imager:

\begin{equation}
	FWHM_{s} \approx \frac{\lambda}{r_{0}}
\end{equation}
where $\lambda$ is the considered wavelength and $r_{0}$ is Fried's coherence length (Fried \shortcite{Fried1966}), at that wavelength:

\begin{equation}
r_{0} = (0.423  k^2 \sec(\zeta) \mu_{0})^{-\frac{3}{5}}
\end{equation}
where $k$ is the wavenumber, $\zeta$ the zenith angle, and $\mu_{n}$ is the n$^\R{th}$ moment of turbulence:

\begin{equation}
\mu_{n}=\int dh C_{n}^{2}(h) h^{n}
\end{equation}
$C_{n}^{2}(h)$ is the refractive index fluctuation structure  coefficient, and
$h$ is the height in the atmosphere.  

Histograms   of    the  seeing in  La    Silla  and  Paranal     are  shown on
Fig~\ref{fig:seeing_histo}.  These  histograms are obtained  at  the two sites
for several years.  Therefore, seasonal variations are smoothed out.

Measurement  of  $C_{n}^{2}(h)$ profiles made by   SCIDAR  or balloon (Sarazin
\shortcite{LASSCA}, Fuchs \& Vernin  \shortcite{PARSCA}) show  that turbulence
is usually concentrated in a few thin layers. Therefore,  we have chose to use
the approximation  of infinitely thin turbulent  layers.  We used three layers
to model  the $C_{n}^{2}(h)$ profile.  One represents  the seeing close to the
ground  (telescope  - mirror   - dome   ),   and  two layers   of  atmospheric
turbulence. The height of these layers were chosen so as to match the measured
profiles.

Two atmosphere  models were considered for each  site, in  order to study what
effects  variability in the $C_{n}^{2}(h)$  profile would have on the results.
Therefore, we considered a good  atmospheric condition  case (obtained 20  per
cent of the time), along with a median case (50 per  cent of the time).  These
models are good and median in two ways: the better case represents good seeing
condition, with a favorable turbulence profile (large isoplanatic angle).  The
median  model  is more pessimistic both   in terms  of  seeing and isoplanatic
angle.   Therefore, with these models,  it should be  possible to span all the
possible conditions that will be encountered.

Wind  profiles  are necessary   to   assess  the temporal  behavior    of  the
atmosphere. For the  Paranal wind profile,  a modified Bufton wind profile was
used  (Bonaccini \shortcite{Bonaccini1996}).  For La Silla,  we used a similar
model  but   the low  altitude  wind-layer was   modified according to balloon
measurements made during the LASSCA campaign.

\begin{equation}
v(h)=a_1+a_2 \cdot e^{-(\frac{h-10000}{h_2})^2}+a_3 \cdot e^{-(\frac{h-5000}{h_3})^2}
\end{equation}
$v(h)$  is in meters per second  and $h$ in meters above  the observatory.  We
used: $a_1 = $ 5 , $a_2 = $ 25, $a_3 = $ 10, $h_2 = $ 2000, $h_3 = $ 500.  For
Paranal: $a_1 = $ 5 , $a_2 = $ 25, $a_3 = $  18, $h_2 = $  2000, $h_3 = $ 500.
Table ~\ref{atm_param} summarizes the atmospheric parameters.

\begin{table}
\caption[]{Atmospheric parameters}
\label{atm_param}
\begin{center}
\begin{tabular}%
{|p{1.8cm}||p{0.9cm}|p{0.9cm}||p{0.9cm}|p{0.9cm}|p{0.9cm}|}
\hline
  & La Silla (good) & La Silla (med.) & Paranal (excel.) &Paranal (good) & Paranal (med.)
\\ \hline
Seeing$^{1}$ (\arcsec) & 0.6 & 0.9 & 0.3 & 0.5 & 0.7
\\ 
Dome~seeing(\arcsec)& 0.8 & 0.8 & 0.0 & 0.0  & 0.0 
\\ 
\% 1$^{st}$ layer $^{2}$ & 0.8 & 0.5  & 0.9 & 0.89 & 0.7
\\ 
\% 2$^{nd}$ layer $^{2}$ & 0.2 & 0.5 & 0.1 & 0.11 & 0.3
\\ 
$H_{1}$ (km)& 1.5  & 3.0 & 2.5 & 2.5  & 2.5 
\\ 
$H_{2}$ (km) & 12.0  & 12.0  & 10.0 & 10.0  & 10.0
\\ 
$\theta_{0}$$^{1}$(\arcsec)& 2.3 & 0.9 & 6.0 & 3.5 & 1.7
\\ 
$h_\R{ao}$ (km) & 2.7 & 5.8 & 3.6 & 3.8 & 5.5
\\ 
$\tau_{0}$$^{1}$ (ms) & 5.0 & 3.1 & 11.4 & 6.6 & 3.0
\\ \hline
\end{tabular}
\end{center}
$^{1}$: at 0.5 $\mu$m, zenith\\
$^{2}$: in \% of the total atmospheric seeing\\
\end{table}

In this table: $h_\R{ao}=(\mu_{5/3}/\mu_{0})^{3/5}$ is the weighted altitude of turbulence for Adaptive Optics ~\cite{Roddier1982}, and $\theta_{0}=(2.91 k^{2} \sec(\zeta)^{8/3} \mu_{5/
3})^{-3/5}$ is the isoplanatic angle (in arc-seconds) (Fried \shortcite{Fried1982}), and  $\tau_{0}$ is the correlation time of the atmosphere (Greenwood \shortcite{Greenwood1977}):
\begin{equation}
\tau_{0}= (2.91 k^2 \sec(\zeta) v_{5/3})^{-3/5}
\end{equation}
where $v_{n}=\int dh C_{n}^{2}(h) v(h)^{n}$ is the $n^\R{th}$ wind moment.

\subsection{The NGS AO simulation}
In the following  section, we concentrate on  the NGS-AO performances  for the
3.6m and 8m telescopes.    This simulation  is  based  on analytical   or semi
analytical formulae.

The AO system  simulation is made of  two main parts: the  tip-tilt correction
loop and the higher order correction path.

Tip-tilt  sensing is performed   with quad-cell avalanche photodiodes  (APDs).
We considered three  error  sources: delay  error, photon  noise and sensor
noise.  We  took a small  (0.1 electrons  rms)  additive noise on  the APDs to
allow for some imperfection.

The delay error is caused  by the lag between the  sensor measurement and  the
applied correction.  We supposed that the lag  is only due to  the integration
time on the sensor.  We compared several expressions for $\sigma_{del_t}^{2}$,
the  variance caused by  the  delay (Olivier \& Gavel \shortcite{Olivier1994},
~\cite{Sandler1994}, Parenti \& Sasiela \shortcite{Parenti1994}).  The results
were very similar.  We chose the last  reference, giving the  most pessimistic
result:

\begin{equation}
\sigma_\R{del_t}^{2}=4.09 \cdot \sec(\zeta) D^{-7/3} v_{-1/3}^{8/15} v_{14/3}^{7/15} \tau_\R{dt}^{2}
\end{equation}
where  $\tau_\R{dt}$  is  the tilt  correction  delay  and $D$  the  telescope
diameter.

Photon noise arises from the quantum  nature of light.  It was also taken from
Parenti \& Sasiela  \shortcite{Parenti1994},  which gives similar   results as
Rousset  \shortcite{Rousset1994}.     It's variance,  $\sigma_{ph_t}  ^2$, was
computed according to:
\begin{equation}
\sigma_\R{ph_t} ^2 = \frac{4}{3} \pi ^2 (\frac{k_\R{t}}{k_\R{sc}})^{12/5} \frac{1}{N_\R{ph_t}}
\end{equation}
where $k_{t}$ and $k_{sc}$ are the wavenumbers  of the tilt sensor and science
camera, respectively,  $T_\R{tilt}$ is the  transmission of the tilt  path and
$\eta_\R{tilt}$ is the quantum efficiency of the tilt sensor, $N_\R{ph_t}$ is the number of photons per sub-aperture and exposure time:

\begin{equation}
N_\R{ph_t} = \frac{\pi ^2}{2 h c k_\R{t}} T_\R{tilt} \eta_\R{tilt}  \tau_\R{dt} D^2 I_\R{t}
\end{equation}
where $I_\R{t}$ is the irradiance (in W/m$^2$) at the tilt sensor.

$\sigma_\R{n_t}^2$, the wavefront variance caused  by sensor additive noise is
from Rousset \shortcite{Rousset1994} which seems to fit  better the low fluxes
reference  stars than Parenti   \&   Sasiela \shortcite{Parenti1994}. It   was
slightly   modified,   using  the method  described    in  Parenti  \& Sasiela
\shortcite{Parenti1994}    to   take  into   account   closed-loop  operation,
contributing a factor 2/3 to the variance:

\begin{equation}
\sigma_\R{n_t}^2 = \frac{128}{9} \pi^2 \frac{1}{k_\R{t}^2} \frac{N_\R{rms}^2} {D^2 N_\R{ph_t}^2} (\frac{k_\R{t}}{k_\R{sc}})^{12/5} (\frac{D}{r_{0}})^2
\end{equation}
where $N_\R{rms}$ is the rms readout noise of the detector.

To get the total noise, we assumed uncorrelated errors:
\begin{equation}
\label{eq:tilt_variance}
\sigma_\R{tilt}^2=\sigma_\R{del_t}^{2} + \sigma_\R{ph_t} ^2 + \sigma_\R{n_t}^2
\end{equation}

The  WFS   was modeled in   a  similar way.   We  used  a   system  based on a
Shack-Hartmann (SH) wave front sensor and a CCD. The error sources that appear
with the WFS will now be described.

The  fitting error  depends only of  the deformable   mirror (DM). It  appears
because the actuator spacing is not infinitely small, so all spatial frequency
aberrations      cannot      be       compensated.      We     used  Greenwood
\shortcite{Greenwood1979} for $\sigma_\R{fit} ^2$, the fitting error variance:

\begin{equation}
\sigma_\R{fit}  ^2  =  0.34  \cdot  (\frac{d_\R{s}}{r_{0}})^\frac{5}{3}
\end{equation}
where $d_\R{s}$ is the size of a square sub-pupil.

Another source of  error associated with  the DM and the  WFS  is the aliasing
error.  It results from  the spectral  aliasing of  the high order  modes into
lower order modes. Therefore, the measurement of the wavefront is biased.  The
expression for  this    variance,  $\sigma_\R{alias}  ^2$,  is  from    Rigaut
\shortcite{Rigaut1996}:

\begin{equation}
\sigma_\R{alias} ^2 = 0.2 \cdot (\frac{d_\R{s}}{r_{0}})^\frac{5}{3}
\end{equation}

The sum of the  aliasing and fitting  error variances fixes the maximum Strehl
that can be achieved with an AO system. We have:

\begin{equation}
\sigma_\R{alias}^2 + \sigma_\R{fit} ^2 = 0.54 \cdot (\frac{d_\R{s}}{r_{0}})^\frac{5}{3}
\end{equation}

This is in agreement with  the 0.5 coefficient  proposed by Parenti \& Sasiela
\shortcite{Parenti1994} and 0.54 ~\cite{Sandler1994}.

There are also sources of error similar to those of the tilt sensor.  Like for
tilt,   sensor noise     $\sigma_\R{n}    ^2$  was  derived     from   Rousset
\shortcite{Rousset1994}:

\begin{eqnarray}
\label{eq:read_noise}
\sigma_\R{n} ^2 &=&  \frac{8}{9}\pi ^2 (\frac{k_\R{sc}}{k_\R{wfs}}) ^{2} \frac{N_\R{rms}^2 N_\R{cen}^4}{N_\R{ph}^2} \nonumber\\
&&  (1 +(\frac{k_\R{wfs}}{k_\R{sc}})^{12/5} (\frac{d_\R{s}}{r_{0}})^2)^2  (\frac{k_\R{sc}}{k_\R{wfs}})^{12/5} (\frac{r_{0}}{d_\R{s}})^2
\end{eqnarray}
where $k_\R{wfs}$ is  the wavenumber  of the wavefront-sensor,  $\eta_\R{wfs}$
its  quantum  efficiency   and  $T_\R{wfs}$   the  transmission   of  the  WFS
path. $\tau_{d}$ is the  integration time  of the WFS   and $N_\R{ph}$ is  the
number of photons per sub-pupil per integration time:

\begin{equation}
N_\R{ph} = \frac{2 \pi}{ h c k_\R{wfs}} T_\R{wfs} \eta_\R{wfs}  \tau_\R{d} d_\R{s}^2 I_\R{wfs}
\end{equation}
where $I_\R{wfs}$ is the irradiance (in W/m$^2$) at the WFS.

We used the photon noise expression given by Parenti \& Sasiela \shortcite{Parenti1994}:

\begin{equation}
\label{eq:ph_noise}
\sigma_\R{ph}^2 = \frac{4 \pi ^2}{3} (\frac{k_\R{sc}}{k_\R{wfs}})^2 \frac{1}{N_\R{ph}}
\end{equation}

The bandwidth error (or  time  delay  error)  $\sigma_\R{del}^2$  is  the same
effect than with tip-tilt time delay, but applied to the high order  modes. We
considered:

\begin{equation}
\sigma_\R{del}^2 = 0.962 (\frac{\tau _{d}}{\tau _{0}})^{\frac{5}{3}}
\end{equation}

The different error sources were assumed to be un-correlated:
\begin{equation}
\label{eq:ho_variance}
\sigma_{ho}^2=\sigma_\R{alias}^2+\sigma_\R{fit}^2+\sigma_\R{del}^{2}+\sigma_\R{ph}^2 + \sigma_\R{n}^2
\end{equation}
The noise of the sky background was taken into account by  replacing $N_\R{rms}$
with an equivalent noise $N_\R{eq}$ according to:
\begin{equation}
N_\R{eq}= \sqrt{N_\R{rms}^2 + N_\R{sky}}
\end{equation}
where $N_\R{sky}$  is the number  of electrons  of noise  coming  from the sky
background (set  to   $m_\R{R}=21.5$).  It is  small  in  the red, where   the
wavefront sensing is done.  However, it becomes non-negligible in the LGS case
with tilt sensing on a natural guide star.  The  sky-background was assumed to
be recorded  before the loop is closed  and therefore only  the sky background
remains. We  converted magnitudes to  fluxes with:
\begin{equation}
I=10^4 \cdot \Delta \lambda \cdot 10^{-(\frac{m}{2.5}-ZP)}
\end{equation}
where I is the irradiance (in W/m$^2$), $m$ is the magnitude, $\Delta \lambda$
is  the bandwidth (in  microns),  and ZP  the magnitude  zero point (-11.75 in
R). Table ~\ref{sys_parameters}  sums up the  instrumental parameters used  in
this study.

\begin{table}
\caption[]{Simulation parameters}
\label{sys_parameters}
\begin{center}
\begin{tabular}%
{|p{3cm}||l|l|}
\hline
  & 3.6-m & 8m
\\ \hline
D (m) & 3.6  & 8.0
\\ 
$d_\R{s}$ (m) (minimum) & 0.5  & 0.5 
\\ 
$\eta$ (WFS-Tilt) & 0.9 - 0.6  & 0.9 - 0.6
\\ 
$\Delta \lambda$ (WFS-Tilt) ($\mu m$)& 0.3 - 0.4  & 0.3 - 0.4
\\ 
$T_\R{wfs}$ - $T_\R{tilt}$ & 0.35 - 0.30 & 0.40 - 0.35
\\ 
$N_\R{rms}$ (WFS-Tilt) & 4.0 - 0.1 & 3.0 - 0.1
\\ \hline
\end{tabular}
\end{center}
\end{table}

Computing the long exposure  Strehl ratio, $\R{S}_\R{le}$, from the  wavefront
variance  is a  non-trivial matter,  when these  variances are high (typically
more than  $3-4 \R{rad}^{2}$).   Parenti \shortcite{Parenti1992}  assumes that
the PSF  consists of two components:  a diffraction limited   core and a halo.
The width of the halo is that of an uncompensated image. In reality, this halo
is   narrower  than the   seeing  disk  (e.g.  ~\cite{Rigaut1997b}),  but  this
approximation is  good enough for  our purpose.   It is more  accurate at high
wavefront   variances than identifying  the   Strehl with the coherent  energy
($\R{S}=e^{-\sigma^2}$).

The FWHM of the PSF can also be estimated, using the same reference:

\begin{equation}
\R{S}_\R{le}=\frac{e^{- \sigma_\R{ho} ^2}}{1+ \frac{\pi ^{2}}{2} (\frac{D}{\lambda_\R{sc}})^{2} \sigma _\R{tilt}^{2}}+ \frac{1-e^{- \sigma _\R{ho}^{2}}}{1+ (\frac{D}{r_{0}})^2}
\end{equation}
\begin{equation}
FWHM= 1.03 \frac{\lambda_\R{sc}}{D} \frac{\{\frac{e^{-2 \sigma _\R{ho} ^2}}{[1+ \frac{\pi ^{2}}{2} (\frac{D}{\lambda_\R{sc}})^{2} \sigma_\R{tilt}^{2}]}+ \frac{(1-e^{- \sigma _\R{fig} ^{2}})^{2}}{(1+\frac{D}{r_{0}})^{2}}\}^{1/2}}{S_\R{le}}
\end{equation}
where $\sigma_\R{ho}$ is the figure error, $\sigma_\R{tilt}$ is the tilt error and  $\lambda_\R{sc}$ the science wavelength.

\begin{figure*}
\centerline{\psfig{figure=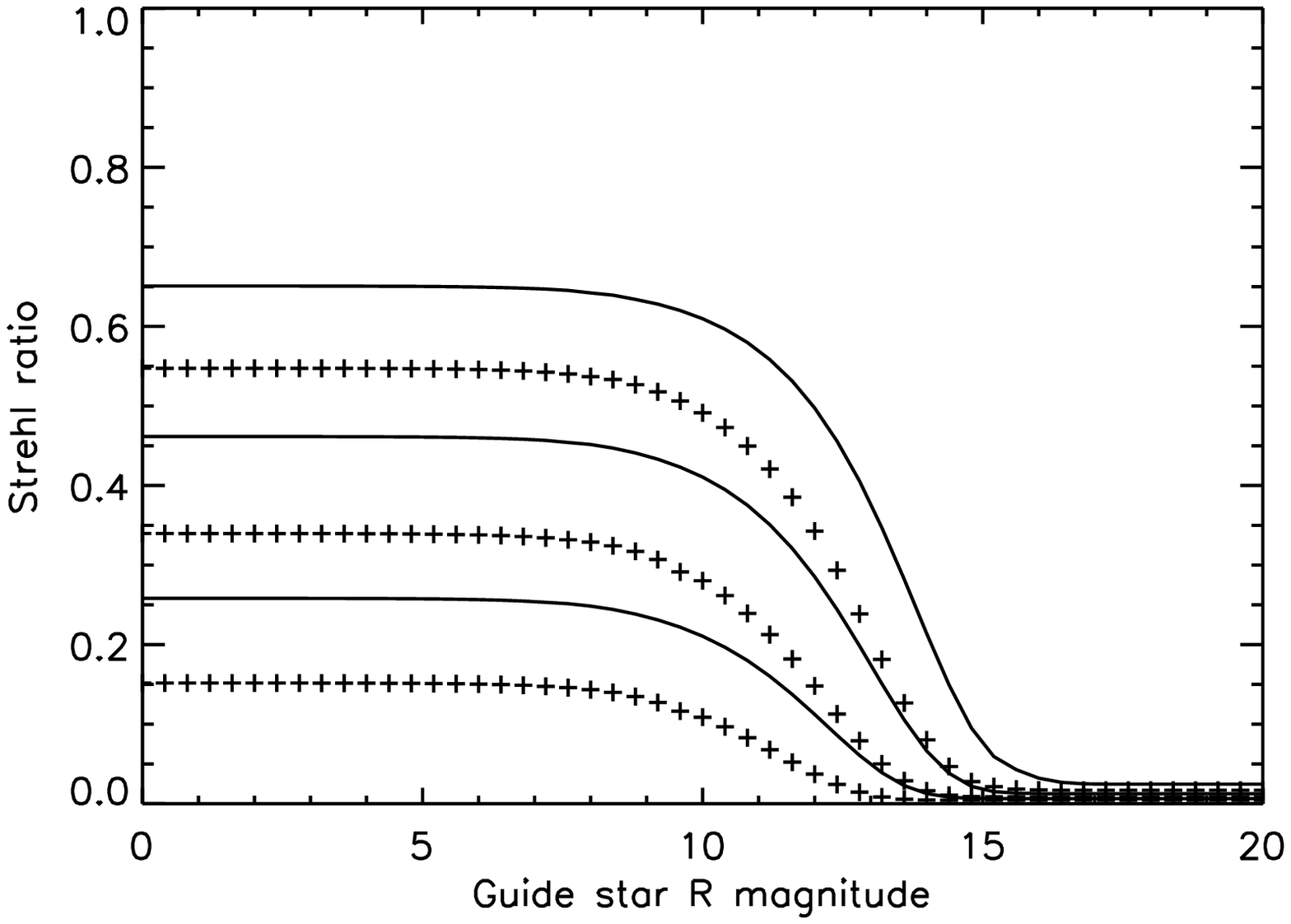,width=9.0cm}\psfig{figure=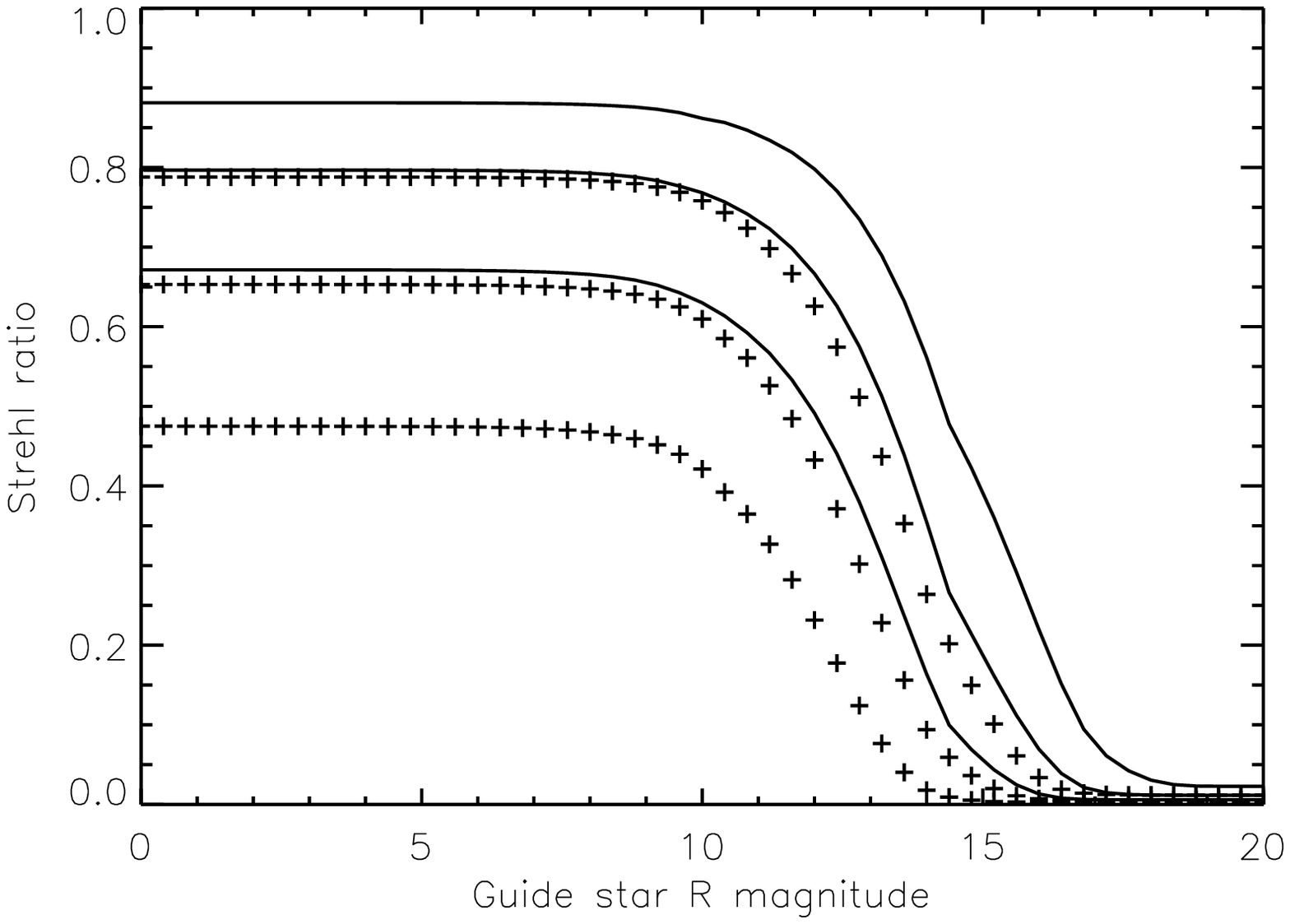,width=9.0cm}}
\caption{Predicted Strehl vs guide star magnitude for 3.6m (left) and 8m(right), at 2.2 $\mu$m, 1.65 $\mu$m, 1.25 $\mu$m, good model (solid) and median model(crosses).}
\label{fig:AO_plots}
\end{figure*}

To get $\tau_{dt}$ and $\tau_{d}$,  the integration times  on the tip-tilt and
wavefront   sensors, we  used a  Powell   optimization algorithm  to find  the
integration times  that yield maximum Strehl.   The maximum sampling frequency
is set by the AO system hardware and therefore was limited to 500 Hz.

Another optimization scheme was to change the size of the sub-pupils (from 0.5
m, 1.0  m,  2.0 m corresponding  on  the 8m telescope to   16x16, 8x8,  4x4 SH
configurations).   This method was applied   for all NGS  results presented in
this article.  It allows to increase by approximately 1 the limiting magnitude
in   the 8m case.    For  LGS  results, it   was    checked that the   optimum
configuration was the one with the smallest possible sub-apertures.

Another source of error is introduced if an off-axis reference source is used,
instead  of the science object.  The  method to compute $\sigma_\R{aniso} ^{2}
$,  the  variance    caused  by  anisoplanatism  is   described    in  Chassat
\shortcite{Chassat1989}:

\begin{equation}
\sigma_\R{aniso} ^{2} (\alpha)=2(C_{nn}(0) - C_{nn}(\alpha))
\label{eq:chassat}
\end{equation}
\begin{equation}
C_{nn}(\alpha)=(\frac{D}{r_{0}})^{\frac{5}{3}} \frac {\int dh C_{n}^{2}(h)S_{n}(\frac{\alpha h}{D/2})}{\int dh C_{n}^{2}(h)}
\end{equation}
\begin{equation}
S_n(x)=3.90 (n+1) \int _{0} ^{\infty} dk k^{-14/3} J_{n+1}^{2} (k) J_{0}(xk)
\end{equation}
where $\alpha$ is the angle between the  reference and the science object, $n$
is the   radial  degree of  the considered   modes, $J_{m}(x)$  is  the Bessel
function of   order $m$. The variances   for anisoplanatism are  then added in
eq.~\ref{eq:tilt_variance}  ($n=1$) and eq.~\ref{eq:ho_variance}   ($n$ = 2 to
10).

\begin{figure}
\centerline{\psfig{figure=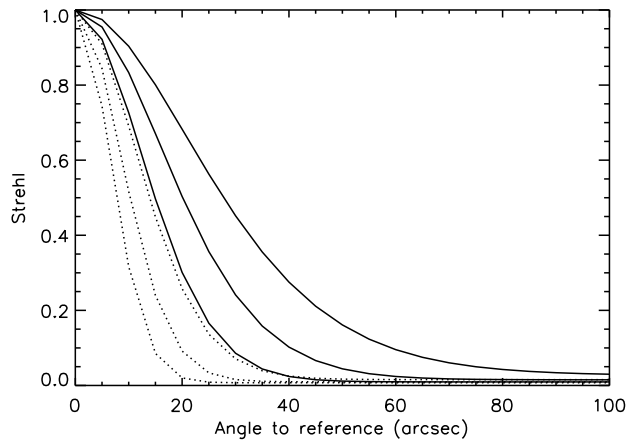,width=9.0cm}}
\caption{Anisoplanatism effect for an 8m telescope at 2.2 $\mu$m (K band), 1.65 $\mu$m (H band), 1.25 $\mu$m (J band), good model (solid) and median model (dots).}
\label{fig:aniso_ho}
\end{figure}

The results   are in  Fig~\ref{fig:aniso_ho}.    At 2.2  $\mu$m,   for a  good
atmospheric model,  a Strehl attenuation  of 50 per cent  occurs at 30 arcsec.
This is a fairly  fast drop, since 80  per-cent of the  Strehl  is lost  at 45
arc-seconds, for the same conditions.

The  Strehl   versus   magnitude of   guide  star magnitude   is   plotted  in
Fig~\ref{fig:AO_plots}. For the 3.6-m  case, peak performance is achieved down
to a guide star magnitude of 10. At $m_\R{R}$ 14, the Strehl is  0.2 in K, for
the good atmospheric model.   It drops to 0.1 with  the median model.  For the
8m case, peak performance  (S=0.86) is also achieved to  $m_\R{R}$ = 10. S=0.2
is  reached at magnitude 16,  for the  good model,  in K.   The improvement in
performances between  the two systems, which  have the same sub-aperture size,
is mainly  due to a  better seeing,  and in lesser  extent  to better hardware
characteristics (e.g.  less  readout noise in  the CCD).  Static  aberrations,
which can  reduce  significantly the  Strehl ratio  have not been   taken into
account in this study. Therefore,  in a real system,  the peak Strehl ratio at
2.2 $\mu$m should be reduced by about 10 per cent with a careful design.

\subsection{AO simulation validation}

The  validity of   this analytic  model  was  checked  with  a  full numerical
simulation  of the  AO   system.    This code,   developed    by F.     Rigaut
\cite{Rigaut1997a} uses    Kolmogorov phase   screens,  simulates  SH  images,
calculates the  centroids and deduces the  DM commands. The  outputs are short
and long exposure PSFs.

\begin{figure}
\centerline{\psfig{figure=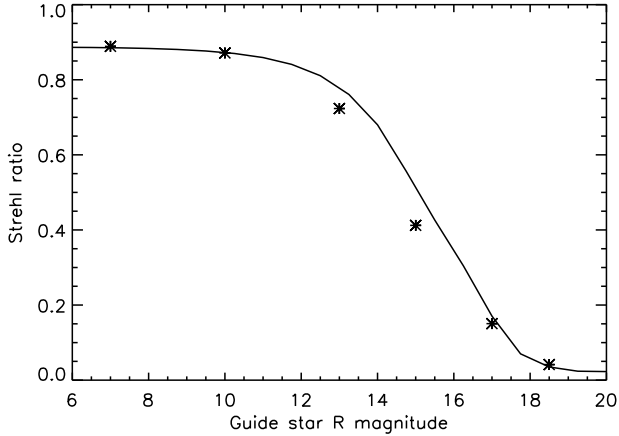,width=9.0cm}}
\caption{Comparison between the analytical AO model (solid) and the numerical model (stars) for 8m telescope at 2.2 $\mu$m.}
\label{fig:comp_anal}
\end{figure}

We simulated the 8m  AO system with  this code.  Integration time, thresholds,
sub-aperture  configuration, loop gain were  optimized  to get the best Strehl
ratio.  Fig~\ref{fig:comp_anal} shows  that  there is good  agreement  between
analytical  and    numerical  models.  However  the  analytical    model gives
systematically better results than the numerical code, which could be expected
since in  the analytical model noise is  not treated  as  precisely as  in the
numerical  model.   For example,  speckles on  sub-images  are  not taken into
account, and noise propagation is less  accurately modeled. The good agreement
at the  magnitude of the   LGS allows us   to use the   analytical model.  The
behavior of the anisoplanatic effect was also validated with this model.

\subsection{The LGS simulation}

In the following, we assume that a sodium laser  guide star, located at 90 km,
is used. The laser power of 3-5  Watts creates a  10.5 magnitude guide star in
the  sodium layer.  These values are  representative of what has been achieved
in experimentally (see e.g. ~\cite{Jacobsen1994}).

Four  effects  specific  to  the  LGS  were  simulated:   cone   effect,  tilt
anisoplanatism, error on the WFS due to  the  spot  size  and  outer  scale of
turbulence (for the 8m case).

The cone effect was first  pointed out by  Foy \& Labeyrie \shortcite{Foy1985}
and has been studied  by several authors (Fried \shortcite{Fried1994}, Sasiela
\shortcite{Sasiela1994}, Tyler \shortcite{Tyler1994}).  The estimations  given
by these authors of the variance of  the wavefront due  to the cone effect are
in   good    agreement.      Therefore,     the  fastest     method     (Tyler
\shortcite{Tyler1994}) was used.  The variance due to the cone effect is:
\begin{equation}
\sigma_\R{cone} ^{2}= (\frac{D}{d_{0}})^{\frac{5}{3}}
\end{equation}
\begin{equation}
d_{0}= \lambda ^{\frac{6}{5}} \cos^{\frac{3}{5}}(\zeta)[\int dh C_{n}^{2}(h) F(\frac{h}{H})]^{-\frac{3}{5}}
\end{equation}
where F is a numerical function and H is the laser star height.

\begin{figure}
\centerline{\psfig{figure=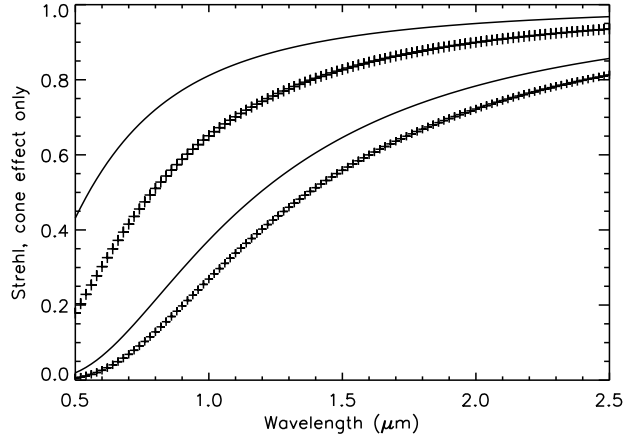,width=9cm}}
\caption{Wavelength dependence of cone effect for a 3.6m telescope (solid) and
an   8m   telescope (crosses). Good  (above)   and  median (below) atmospheric
models}\label{fig:cone}
\end{figure}

Figure ~\ref{fig:cone}  represents the loss   of Strehl only  due to  the cone
effect.   On a large   telescope, the cone   effect  rapidly becomes the  most
important source of error.  It  reduces the maximum  Strehl  by a factor of  2
around 1.0  $\mu$m (depending on the $C_n^2$  profile).  A loss of  80 percent
occurs in the red,  between 0.5 and  0.8 $\mu$m.   These curves  underline the
absolute necessity to  find  a solution to  the  cone effect, for  example  by
considering  a   multiple   LGS   scheme     (see   e.g.   Tallon   \&     Foy
\shortcite{Tallon1990}).

\begin{figure}
\centerline{\psfig{figure=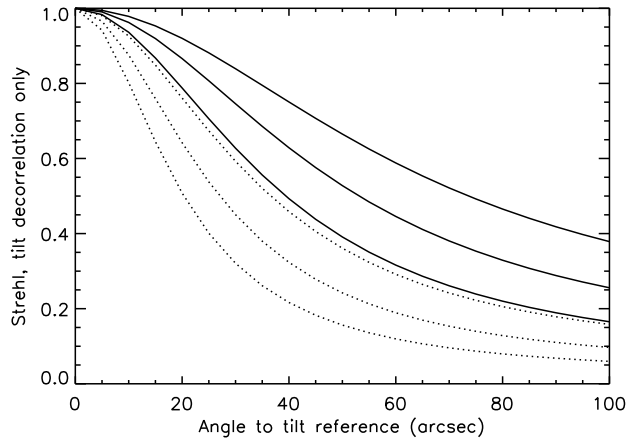,width=9.0cm}}
\caption{Tilt anisoplanatism: Wavelength and model comparisons for an 8m class telescope. Good (solid) and median atmosphere (dots), $\lambda = 2.2 \mu m, 1.65 \mu m, 1.25 \mu m$ (top to bottom)}
\label{fig:tilt_aniso}
\end{figure}

When the science object is not bright enough for tilt sensing,  one has to use
a  nearby  tilt   reference star.    The  variance   caused  by   this effect,
$\sigma_\R{t-aniso}  ^{2}$,   was   computed  with  eq.\ref{eq:chassat}.   The
corresponding loss of Strehl can be seen  on Fig~\ref{fig:tilt_aniso}.  A loss
of 50  per cent occurs,  in K (good  model), at 75 arc-seconds.  These results
are  in good  agreement with  Olivier \&  Gavel \shortcite{Olivier1994} taking
into  account the differences in  turbulence profiles, telescope diameters and
hardware parameters.

The effect of the outer  scale of turbulence  on tilt anisoplanatism could  be
significative, if   the outer scale, $L_{0}$  is  comparable to  the telescope
diameter.  $L_{0}$ is poorly known, the estimates range from a few meters to a
few thousands of meters (see to summary of measurements in ~\cite{Agabi1995}).
However, recent measurements seem to converge to the  same order of magnitude,
from 20  to 50  meters.    Therefore,  we used   the computations  by  Sasiela
\shortcite{Sasiela1994}     to  get the   effects   of   outer  scale on  tilt
anisoplanatism,  in  the 8  meter case, assuming  a  Von Karman spectrum.

\begin{figure}
\centerline{\psfig{figure=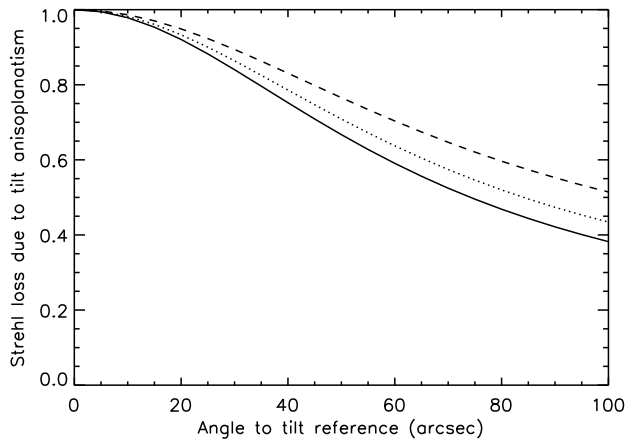,width=9.0cm}}
\caption{Outer scale of turbulence and tilt anisoplanatism: effects on an 8meter class telescope. $L_0= \infty$ (solid), 50 m (dash), 26 m (dot). K band, good model.}\label{fig:outer_scale}
\end{figure}

Three values for $L_{0}$ were  used: 26m, 50m and  infinite. Smaller values of
$L_{0}$ were not considered    because  the expressions derived by     Sasiela
\shortcite{Sasiela1994}   assume   that     $\pi D  \slash     L_{0}    <  1$.
Fig~\ref{fig:outer_scale}  shows that  the effect  of  the outer scale  is not
significant unless it is of the order of the size of the telescope (i.e the 26
meters case). Even in that case,  the variation is  smaller than the variation
caused by seeing  effects (at 100 arc-seconds,  the difference in  the extreme
models is 30 per cent).  Therefore, in the rest of this  article, we assume an
infinite outer scale.

When using  a  LGS,    one has  to   be able   to filter  out     the Rayleigh
backscattering,   created by  the lower  layers   of the atmosphere, which can
contribute to   WFS  noise if  not filtered.   The  rejection can  be  done by
emitting  the  laser through an   off-axis  beam projector   and by  spatially
filtering the Rayleigh light  with a field stop at  the entrance focus  of the
WFS.  In  the  following, the residual   noise due to  Rayleigh scattering was
neglected.

The angle  between the Rayleigh and the  Sodium stars,$\theta$, is  given by a
simple geometrical approach:

\begin{equation}
\theta=\frac{(H_\R{Na}-H_\R{R})d}{H_\R{Na} H_\R{R}}
\label{eq:spot}
\end{equation}
where $H_\R{Na}$ is  the altitude of the  Sodium layer ($\approx$ 90 km),
$H_\R{R}$ is the altitude at the tip of the Rayleigh scattering region and $d$
is the separation of the beam-projector and the  considered sub-aperture.  For
an 8m telescope, assuming a distance of 1m between  the beam projector and the
primary mirror's edge, the angular separation  between the Rayleigh star's tip
and the sodium star  varies from  8 arcsec  to 72 arcsec,  if  the tip  of the
Rayleigh   region   is  20  km.   However,     the ATLAS experiment   (Laurent
\shortcite{Laurent1996}) showed  that at  20 km,  one  can still  measure  the
Rayleigh backscattered  flux.  This is  confirmed by LIDAR studies (see review
by Gardner \shortcite{Gardner1989}), showing that  photocounts due to Rayleigh
scattering reach the level  of background noise near 50  km.  If one considers
that Rayleigh scattering stops at that height, the separation ranges then from
1.8 arcsec to 16.5  arcsec.  This may be  close enough to produce a measurable
effect on the WFS and therefore reduce the quality  of the correction. Another
solution is to propagate the beam from the back of  the secondary mirror.  The
separation  of the  two stars  would be done  with  the  help  of the  central
obscuration (The Rayleigh scattering would be hidden by the secondary).

The elongation of  the laser sodium spot,  due to off-axis propagation, can be
studied with a similar formula:
\begin{equation}
\theta_\R{spot} \sim \frac{\Delta H d }{H^2}
\end{equation}
where $\Delta H$  is the thickness of  the sodium layer (10  km). We obtain an
elongation ranging from 0.25 arcsec to 2.3 arcsec.   

The spot size  can also be enlarged  by the limitations of  the optics and the
laser itself. Therefore,  we assumed that the  spot size is independent of the
seeing and fixed its value to 1.5 arcsec, with the  elongation of the spot and
with measured values of  spot   size (see  e.g.  ~\cite{Jacobsen1994}).    The
increased size reduces the signal to noise ratio on the WFS.  We neglected the
effect of varying spot size on the pupil and took the mean value of elongation
(center of the   pupil to propagator).  An   effective  $r_{0}$ was  computed,
giving the desired   spot size.  This  mean size   was injected in  the  noise
calculations    of     the      WFS    (in   eq.~\ref{eq:read_noise}       and
eq.~\ref{eq:ph_noise}).  The effects of this spot size are  small in K band (5
per   cent maximum  Strehl   reduction),  but  become  noticeable  at  shorter
wavelengths (30 per cent in the I band).

Measurements of    the height, density  and   thickness  of the   sodium layer
(~\cite{Papen1996}) show significant    variations in  these quantities   over
periods of minutes.   These variations can  change the LGS brightness, and the
variation of  the focus. Therefore,  sensing of the  focus from another source
than the LGS should be further investigated.
\section{Results}
\subsection{Estimated system performance}

\begin{figure*}
\centerline{\psfig{figure=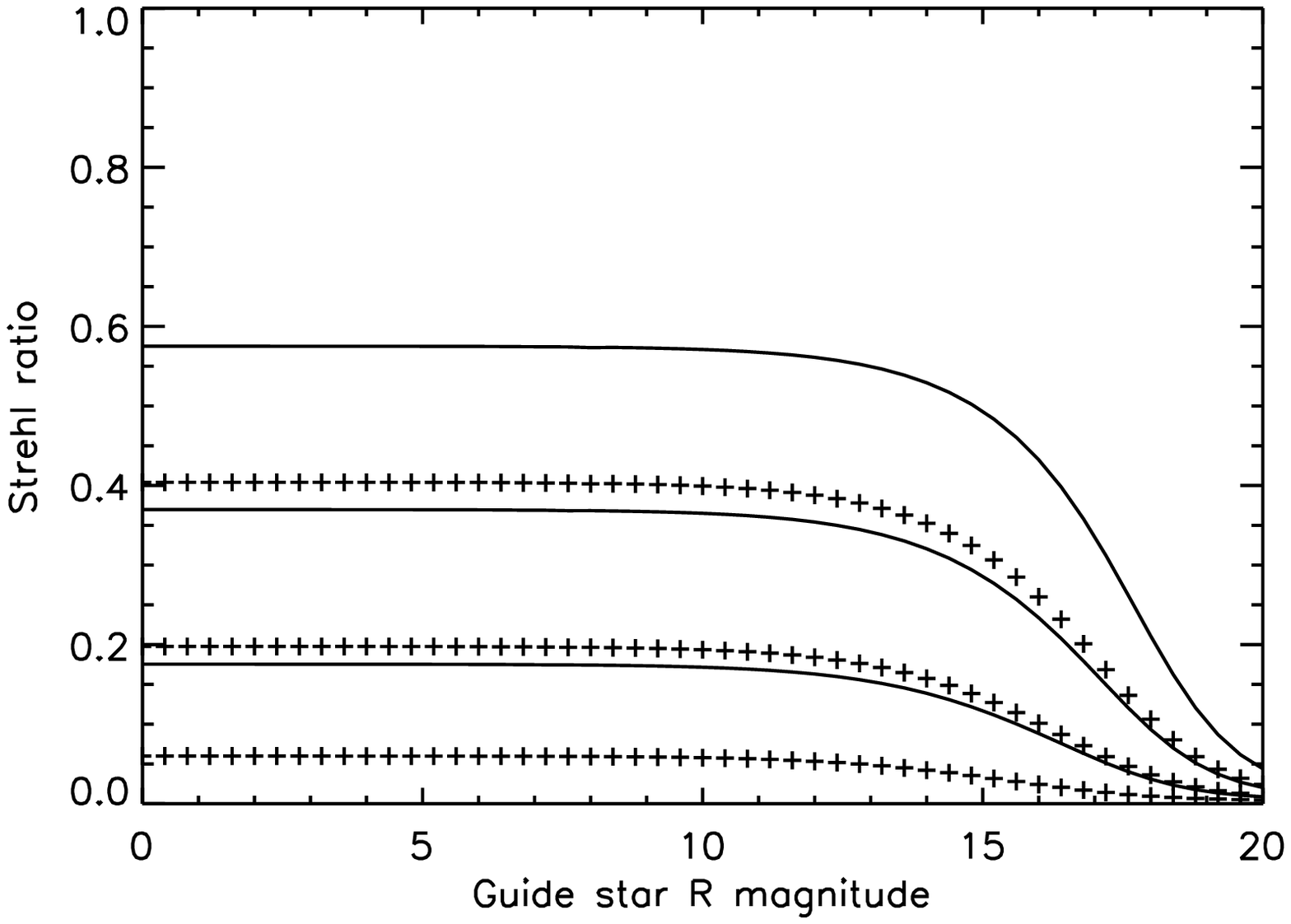,width=9.0cm}\psfig{figure=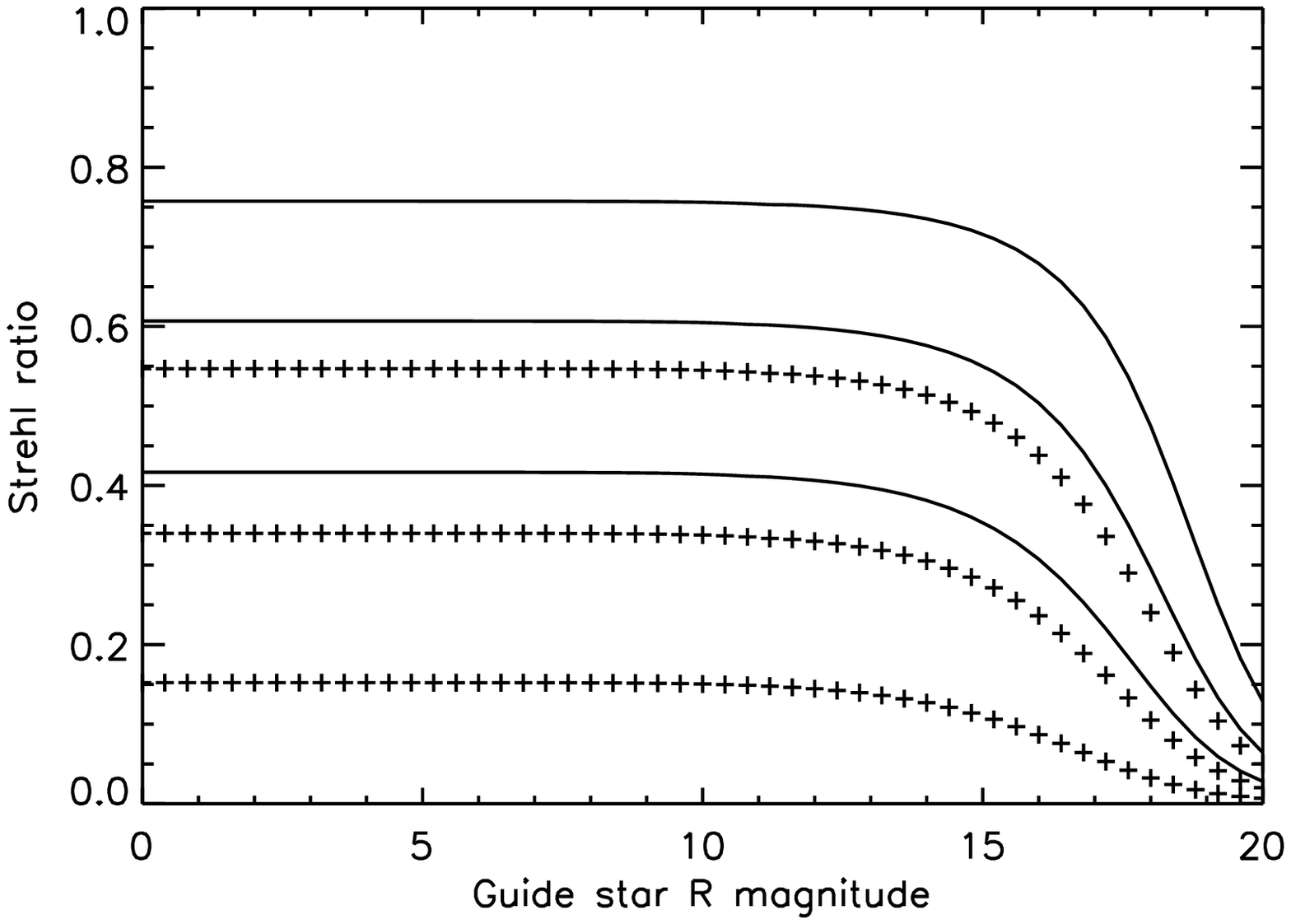,width=9.0cm}}
\caption{Performances for the AO+LGS systems (left: 3.6m, right: 8m), good (solid) and median (crosses) atmospheric model, in K, H, J bands (top to bottom). The reference star is on-axis.}
\label{fig:ao_lgs}
\end{figure*}

The final  result for the Laser  guide star system is  obtained by adding the
LGS-specific variances, neglecting possible correlations:

\begin{equation}
\sigma ^{2}_\R{tilt-LGS} = \sigma ^{2}_\R{tilt} + \sigma ^{2}_\R{t-aniso}
\end{equation}
\begin{equation}
\sigma ^{2}_\R{ho-LGS} = \sigma ^{2}_\R{ho} + \sigma ^{2}_\R{cone}
\end{equation}

The  results   showing the performances   of  a laser   guide  star system are
presented on Fig~\ref{fig:ao_lgs} where it was assumed that the tilt reference
is  on-axis (i.e.   the science  object is bright   enough for tilt  sensing).
Comparing  the Strehl vs   magnitude plots obtained in Fig.~\ref{fig:AO_plots}
and  Fig.~\ref{fig:ao_lgs}, we  can  see that  the laser   star  has two  main
effects.  It lowers the maximum Strehl that can be achieved with the AO system
and drops, for the 8m system, from  0.87 to 0.75 in K  for the good atmosphere
(14 per cent reduction).  At shorter  wavelengths and median atmosphere model,
the Strehl reduction is  more severe (60 per cent  in J for the median model).
The effect is less severe on a 3.6m telescope  (maximum loss: 60 per cent, the
difference between the  two La Silla seeing  models being larger, as  shown by
Fig.~\ref{fig:cone}).  The second  effect  is that the limiting  magnitude  is
much fainter with an LGS system.  The 8m LGS-based  system reaches a Strehl of
0.2  (K, good model)  down to magnitude  19.5, which is 3.5 magnitudes fainter
than with a NGS.  A comparable gain is achieved on the 3.6m.

The cone effect is heavily  $C_n^2$  dependent.  In  good conditions, one  can
achieve approximately the same Strehl in H band that can be achieved in K band
when the atmosphere  is median.  In order to  be able to  use the LGS  when it
brings  the best improvement, the  location of  the dominant turbulence layers
should be known.  If the turbulence is  very high and the  seeing is poor, the
efficiency of the  LGS will be severely affected  (more than with conventional
AO).  This is especially true for  the 8-m case, for which  the cone effect is
the dominant  source  of error.  Therefore a   device measuring the turbulence
profile      in real   time,  for  example     a   SCIDAR  (Azouit  \&  Vernin
\shortcite{azouit1980}) or a more simple  device such as  a measurement of the
isoplanatic   angle ~\cite{Krause-Polstorff1993}, even  with  a  low accuracy,
would help predicting what kind of performances can be  expected from the LGS.
This demonstrates   the importance of  queue  scheduling, which sets observing
priorities according  to the  atmospheric conditions: high  angular resolution
programs should have a high priority when the seeing-conditions are good.

\subsection{FWHM estimation from Strehl}

\begin{figure*}
\centerline{\psfig{figure=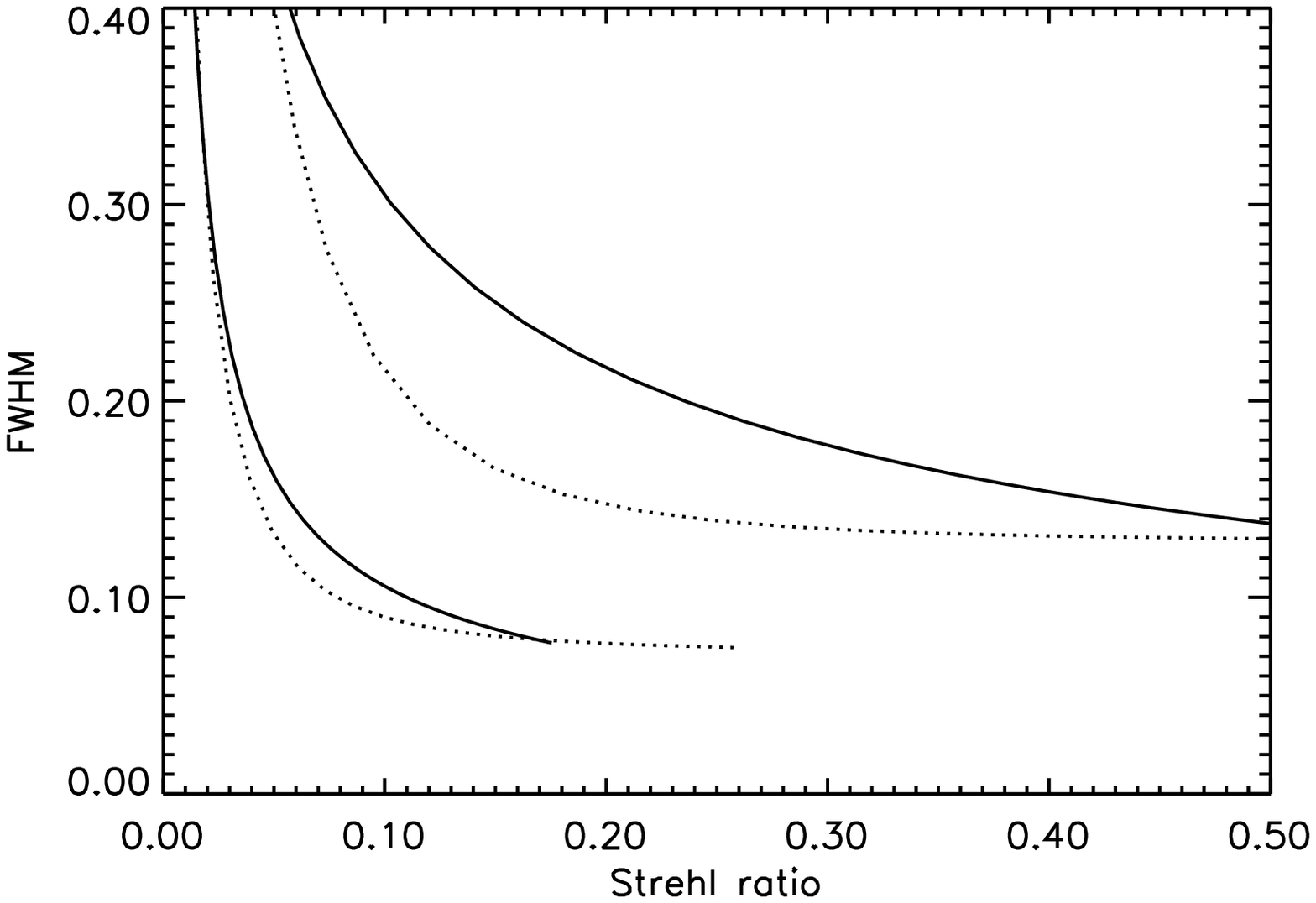,width=9.0cm}\psfig{figure=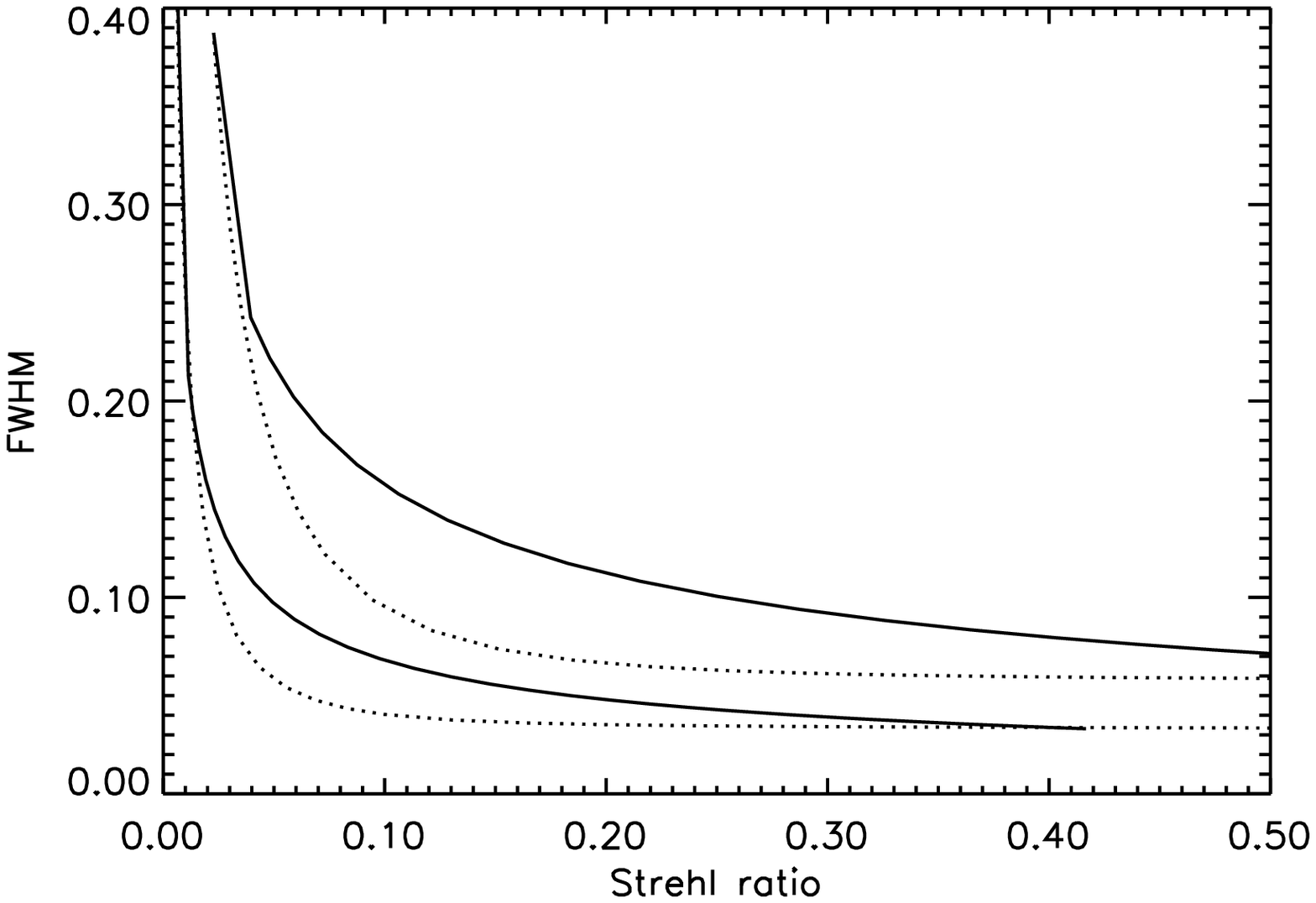,width=9.0cm}}
\caption{FWHM (arcsec) vs Strehl at 2.2 $\mu$m (top) and 1.25 $\mu$m (bottom) for the 3.6 m case (left) and the 8 m case (right), good model. NGS AO (dotted) and LGS AO (solid).}
\label{fig:fwhm_strehl}
\end{figure*}

Usually, astronomers  prefer to use FWHM  of the PSF  as an indicator of image
quality rather than Strehl.  In order to convert one  to the other, we plotted
the FWHM    as a   function    of  Strehl   for     the 3.6m  and     the   8m
(Fig.~\ref{fig:fwhm_strehl}).   These figures    were  done  using  the   good
atmospheric model, varying the magnitude of the natural reference star.  For a
given  Strehl, the LGS  system has a wider PSF  than the  NGS system.  For LGS
systems,  the short-exposure PSF is nearly   diffraction limited , because the
high    orders  are well   corrected by  the   LGS. The   long exposure Strehl
degradation is due to the jitter caused by tilt measurement  error (due to the
faint magnitude of the  tilt reference star  and / or by tilt anisoplanatism),
which moves  this coherent peak.   However, the  energy stays  in the coherent
peak, which  is wider because of the  jitter, but does  not  contribute to the
halo, like it does in  a NGS system.  This  should be an advantage for example
in the  detection   of  faint companions.   On   NGS  systems,   the telescope
diffraction  limit  can be achieved for  Strehls  of 0.1  - 0.2.   Below these
values,  the coherent peak still exists  but it is   more difficult to extract
information in these low   corrections regimes. The limit-Strehl below   which
astronomical results cannot be obtained deserves further investigations.

\subsection{Sky coverage}
Two different methods were used to study the sky coverage.  The first approach
is statistical. The second one is a cross correlation of catalogues.
\subsubsection{Statistical approach}

To compute the statistical sky coverage,  we used the  so-called ``Besan\c con
model'' (Robin \& Cr\'ez\'e \shortcite{Robin1986}), which is a synthetic model
of the  Galaxy.  At a given  galactic latitude and  longitude, it provides the
density of stars that can be observed in a wavelength band.

Assuming that the position of  the  stars follow  Poisson statistics, one  can
compute the probability  $P$ to find at  least one star  within a given radius
$r$:
\begin{equation}
 P_{N_\R{stars} > 0}(m, r)=1-e^{- \frac{\pi r^{2} \eta (m)}{3600^{2}}}
\end{equation}
where $\eta(m)$ is  the  density of  stars brighter  than magnitude  $m$  (per
square degree) in the considered region (Besan\c con model).

\begin{figure*}
\centerline{\psfig{figure=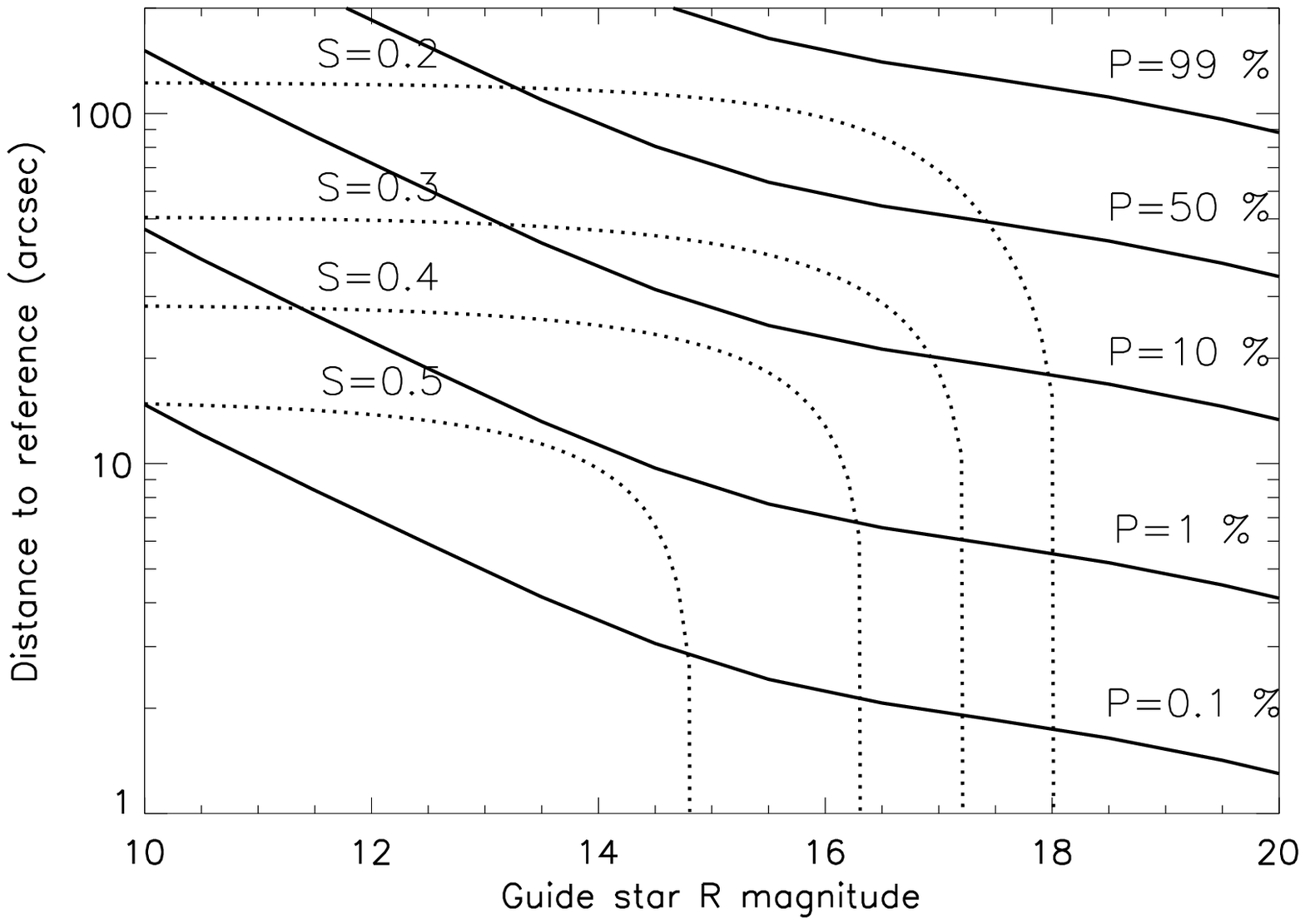,width=9.0cm}\psfig{figure=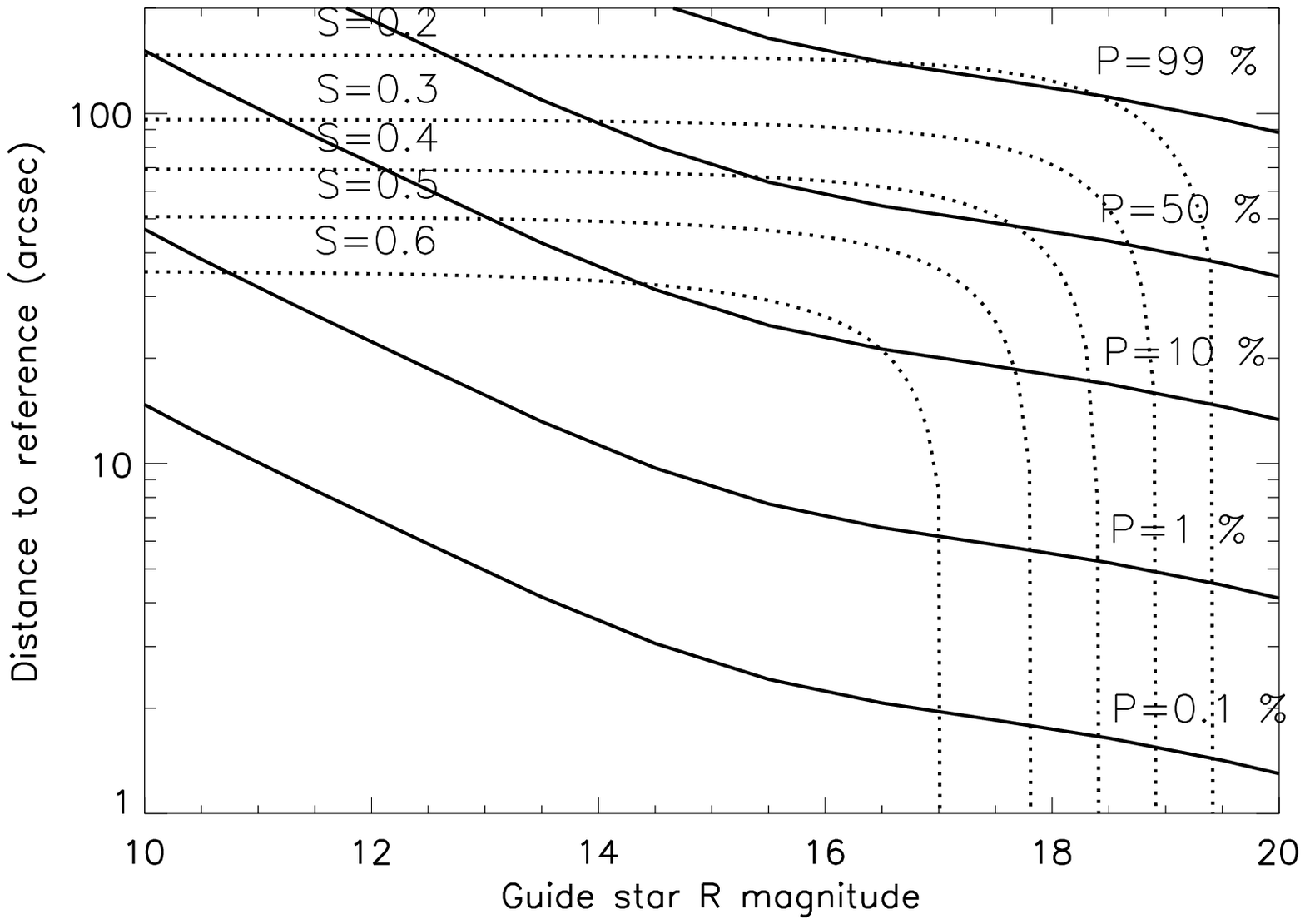,width=9.0cm}}
\caption{Iso-Strehl curves (dotted) for 3.6m + LGS(left) and 8m + LGS (right). Probability to find a reference star (solid) at galactic longitude (l) = 180, latitude (b) = 20. Notice logarithmic scale for distances.}
\label{fig:iso_strehl}
\end{figure*}

For the galactic latitudes  and  longitudes we  considered, we computed  their
most favorable zenith angles.  An  iso-Strehl plot was made  at each of  these
angles.    These plots  were  then  overlaid  on  the iso-probability   curves
(Fig~\ref{fig:iso_strehl}). It allowed us to get  the statistical sky coverage
values.   For the 3.6m  case,  a  K-Strehl  of 0.2  can  be   achieved with  a
probability between 50 and 99  percent. A finer  plot shows the probability to
be 75 per cent.

The results obtained for the 3.6 m telescope and for the 8m in  the K band can
be    found      respectively    on       Tab.~\ref{adonis_results}        and
Tab.~\ref{naos_results}.      Results  for    J     band are     presented  in
Tab.~\ref{adonis_results_J} (3.6m telescope) and Tab.~\ref{naos_results_J} (8m
case).  The first  percentage represents the  probability (in per cent)  for a
given Strehl for the good seeing atmospheric model.  The  second number is for
the median atmospheric model. The Strehl differences between the galactic pole
and the galactic  center can easily be explained.   There is a higher  stellar
density  near the  center,   which allows to   find  a closer reference  star.
However, there is  a second effect that reduces  even  more the Strehl at  the
pole.  Indeed, at  Paranal and La  Silla, the galactic  center is  almost at 0
zenith angle.  On the  other hand, the galactic  pole region is fairly low  in
the sky.  This effect reduces even more the sky coverage obtained.

In K band, the LGS brings a significant improvement of the sky coverage in all
cases,  whether  the atmospheric   conditions are good   or  median and at all
stellar densities.  A  30 per cent  sky coverage (at  0.5 Strehl) is  achieved
with  average stellar densities,  when  a  coverage of only    2 per cent  was
possible with NGS, under good seeing conditions (8m  case).  Note that the sky
coverage  can drop from  80 per cent to  15  per cent  on  a NGS system simply
because the atmospheric conditions change.   For the 3.6  meter case, the gain
goes from a factor of 2 (low Strehl, high star count, K  band), to a factor of
100 at low Strehls and low star counts.

\begin{table}
\caption[]{statistical sky coverage - 3.6m case, K band}
\label{adonis_results}
\begin{center}
\begin{tabular}%
{|p{1.6cm}||p{1.4cm}|p{1.4cm}|p{1.8cm}|}
\hline
K~Strehl & Center & Average$^{1}$ & Pole 
\\ \hline
$S_\R{NGS} = 0.2$ & 20 - 1 & 1 - 0.08 & 0.01 - $<$0.01
\\ 
$S_\R{LGS} = 0.2$ & 99 - 70 & 75 - 5 & 0.5 - $<$0.01
\\ 
$S_\R{NGS} = 0.3$ & 8  - 0.3 & 0.6 -  0.05 & $<$0.01
\\ 
$S_\R{LGS} = 0.3$ & 99 - 10 & 25 - 0.5 & 0.08 - $<$0.01
\\ \hline
\end{tabular}
\end{center}
$^{1}$: l=180, b=20.
\end{table}

For J band, the  conclusion is less obvious. Sky  coverage is low (except near
the pole, where it can be good at low Strehls), and sometimes NGS gives better
performances than  LGS (8m case,  median   seeing, Strehls  of 0.2  and  0.3).
However,  in good seeing cases,  the LGS still   gives better results than NGS
(factor of 30 better for S=0.2 at average densities).  It can go from a factor
of 3 (Strehl  of   0.2) to  more  than  a  factor 100  (near the   pole).  The
sensitivity to seeing  conditions is also greater than  in K band (because
of the cone effect  which is wavelength dependent).  Therefore, in J band, to
get a gain with the LGS, favorable seeing conditions are needed.

\begin{table}
\caption[]{statistical sky coverage - 3.6m case, J band}
\label{adonis_results_J}
\begin{center}
\begin{tabular}%
{|p{1.6cm}||p{1.4cm}|p{1.4cm}|p{1.8cm}|}
\hline
J~Strehl & Center & Average$^{1}$ & Pole 
\\ \hline
$S_\R{NGS} = 0.1$ & 0.8 - $<$0.01 & 0.6 - $<$0.01 & $<$0.01
\\ 
$S_\R{LGS} = 0.1$ & 40 - $<$0.01 & 3 - $<$0.01 & $<$0.01
\\ \hline
\end{tabular}
\end{center}
$^{1}$: l=180, b=20.
\end{table}

\begin{table}
\caption[]{statistical sky coverage - 8 m case, K band}
\label{naos_results}
\begin{center}
\begin{tabular}%
{|p{1.6cm}||p{1.6cm}|p{1.6cm}|p{1.8cm}|}
\hline
K~Strehl & Center & Average$^{1}$ & Pole 
\\ \hline
$S_\R{NGS} = 0.2$ & 80  - 15 & 10  - 1 & 0.5 - 0.05
\\ 
$S_\R{LGS} = 0.2$ & 99 - 99 & 99 - 33 & 10 - 0.8 
\\ 
$S_\R{NGS} = 0.3$ & 60  - 8 & 8  - 0.7 & 0.2 - 0.02
\\ 
$S_\R{LGS} = 0.3$ & 99  - 99  & 75  - 12 & 5 - 0.08
\\ 
$S_\R{NGS} = 0.5$ & 20  - 1 & 2  - 0.1 & 0.07 - $<$ 0.01
\\ 
$S_\R{LGS} = 0.5$ & 99  - 8 & 30  - 0.5 & 0.7  - $<$0.01
\\ \hline
\end{tabular}
\end{center}
$^{1}$: l=180, b=20.
\end{table}

\begin{table}
\caption[]{statistical sky coverage - 8 m case, J band}
\label{naos_results_J}
\begin{center}
\begin{tabular}%
{|p{1.6cm}||p{1.6cm}|p{1.6cm}|p{1.8cm}|}
\hline
J~Strehl & Center & Average$^{1}$ & Pole 
\\ \hline
$S_\R{NGS} = 0.1$ & 30 - 2 & 2 - 0.2 & 0.07 - $<$0.01
\\ 
$S_\R{LGS} = 0.1$ & 99 - 20 & 70 - 1 & 1 - $<$0.01
\\ 
$S_\R{NGS} = 0.2$ & 10 - 0.5 & 0.8 - 0.07 & 0.03 - $<$0.01
\\ 
$S_\R{LGS} = 0.2$ & 99 - $<$0.01 & 25 - $<$0.01 & 0.08 - $<$0.01 
\\ 
$S_\R{NGS} = 0.3$ & 5 - 0.08 & 0.5 - 0.02 & $<$0.01
\\ 
$S_\R{LGS} = 0.3$ & 70 - $<$0.01 & 6 - $<$0.01 & $<$0.01
\\ \hline
\end{tabular}
\end{center}
$^{1}$: l=180, b=20.
\end{table}

\subsubsection{Catalogue cross correlations}
The second approach to study sky coverage is  to  make  cross  correlations of
catalogues. We selected several categories  of  astronomical  objects,  in two
categories: extragalactic and stellar.

In the extragalactic domain, the  Veron-Cetty 96 (Veron-Cetty \& Veron
\shortcite{Veron-Cetty1996})  catalogue  was used.     This  catalogue
contains  8609 Quasars and 2833 Active  Galactic Nuclei (AGN).  In the
stellar domain, the SIMBAD database  was used.  Several types of stars
were studied: Miras (total   number  of objects: 4279), Semi   Regular
pulsating  variables (SRs, 2182  objects), and Pre-Main Sequence stars
(PMS, 928 objects). We used  the US Naval Observatory A-V1.0 catalogue
(USNOC) to search for reference stars around these objects.

Because the  chosen  stellar objects are variable  and  the amplitude of their
variations is not  always well known, an approximation  was used: in  order to
get the brightness of the minimum, 4 magnitudes were added to the magnitude of
the bright phase, found in the catalogue (except for  SRs, 2 magnitudes).  For
Miras,  for example, the  variability ranges  from  2.5  to  6  mag in V  band
~\cite{VanBelle1996},   so  4 magnitudes   can   be  seen as  an   ``average''
variability. Therefore, each variable star has  two magnitudes: the bright and
the faint phase.

Another magnitude correction   was applied.   Usually, the  SIMBAD   catalogue
contain only V-band magnitudes.  However,  the WFS is used in  the red part of
the spectrum  (near the R band).   We  took, as  a  first order approximation,
average  V-R correction terms.  Using the General  Catalogue of Variable Stars
(GCVS, ~\cite{Kholopov1988}), the mean spectral type of M-type Miras (the most
numerous) was computed.   A sample of 400  Miras was used.  The mean  spectral
type is M4.4 at maximum  and M7.1 at minimum.   The corresponding V-R  indexes
are 1.7 and 2.2 respectively. The same procedure was applied to SRs. We used a
sample  of 107 objects,  the spectral type varies  from M4.5 (V-R=1.7) to M6.5
(V-R=1.9).  The  variation is smaller than  for the Miras, which was expected.
For the PMS objects,  we used the  mean  of a sample  of  76 T-Tauri  star V-R
measurements, taken  from  ~\cite{Herbst1994}. This leads to  a   V-R index of
0.79.  Because the spectra of Quasars and AGN is very object-dependent, we did
not apply any correction.   A slight bias  for these objects is therefore  not
excluded. We did  not apply any magnitude correction   to the USNOC  reference
stars, because it contains R-band magnitudes.

One  problem  in  the correlation   approach  is   not  to count   an   object
twice. Indeed, because of the imprecision in the coordinates of an object, one
can select it as a science object, and find it also in the  USNOC and count it
as its own reference.  This would overestimate  the number of objects that can
be observed.  It was decided not to select references closer than 3 arc-seconds
to  the object.  We  do not exclude that  despite this procedure, some objects
have been counted twice.

For computational reasons, a radius limited to  240 arcsec was searched around
the object to find a reference .  The number of references  was limited to the
20 closest.  When a reference was found,  the Strehl was computed, taking into
account  the following  factors:   distance and  magnitude  of  the reference,
brightness  of the  laser  star,  zenith angle  at  Paranal  (La Silla).   The
procedure was applied to NGS   and LGS AO,   for good and median   atmospheric
conditions.  When  several  references  were found  for  an  object,   the one
providing the best Strehl was selected.   Strehl ratios were also computed for
all the  science  objects on-axis, assuming  that  the reference was   a point
source   of  the magnitude stated  in   the catalogue (which  can   be a crude
approximation for AGN for example, which are extended  objects, and leads a an
overestimation of the Strehl  for these objects).   This procedure was applied
in J and K bands.  The results for the 3.6m are in Tab.~\ref{adonis_astro} and
Tab.~\ref{adonis_astro_J}, and  those for the  8m are in Tab.~\ref{naos_astro}
and Tab.~\ref{naos_astro_J}.  We  have  included in these  tables  $<S_{NGS}>$
($<S_{LGS}>$) the mean  Strehl, in per  cent, with NGS (LGS)  on each class of
objects (for variable stars,  the  first number indicates  maximum brightness,
the  second the  minimum  brightness), $\sigma_{S_{NGS}}$ ($\sigma_{S_{LGS}}$)
the standard  deviation on this  Strehl, and $N_{S>0.1}$ and  $N_{S>0.2}$, the
number of objects than can be observed with a Strehl greater than 0.1 or 0.2.

\begin{table*}
\caption[]{3.6m - astrophysical targets, K band. GS: Good seeing, MS: Median seeing. Strehls are in per cent.}
\label{adonis_astro}
\begin{center}
\begin{tabular}{|l||c|c|c|c||c|c|c|c|}
\hline 
  Object  &  $<S_{NGS}>$  & $\sigma_{S_{NGS}}$ &  $N_{S>0.1}$ & $N_{S>0.2}$ & $<S_{LGS}>$ & $\sigma_{S_{LGS}}$ & $N_{S>0.1}$ & $N_{S>0.2}$ \\
\hline 
 Quasar - GS & 2.5 & 2.2 & 115 & 23 & 18.6 & 10.6 & 6352 & 3809\\
 Quasar - MS & 1.4 & 0.6 & 4 & 0 & 7.2 & 5.4 & 1890 & 287\\

 AGN - GS & 6.6 & 11.9 & 426 & 273 & 23.9 & 16.9 & 1931 & 1517\\
 AGN - MS & 3.0 & 6.3 & 206 & 106 & 12.0 & 11.1 & 1198 & 671\\

 PMS - GS & 19.7 - 7.5 & 17.4 - 10.6 & 542 - 185 & 377 - 118 & 40.6 - 28.3 & 12.0 - 13.5 & 900 - 835 & 846 - 671\\
 PMS - MS & 10.7 - 2.8 & 13.0 - 5.1 & 319 - 61 & 195 - 21 & 23.7 - 13.2 & 9.59 - 8.8 & 810 - 521 & 680 - 216\\

 SR - GS & 29.6 - 21.4 & 19.2 - 19.2 & 1686 - 1295 & 1420 - 941 & 35.6 - 34.2 & 16.1 - 15.7 & 1885 - 1891 & 1754 - 1724 \\
 SR - MS & 19.1 - 12.3 & 16.0 - 15.2 & 1339 - 780 & 904 - 546 & 20.5 - 19.0 & 11.7 - 11.2 & 1649 - 1594 & 1232 - 1086 \\ 

 Miras - GS & 30.8 - 13.6 & 18.4 - 16.2 & 3526 - 1697 & 2944 - 1115 & 35.8 - 31.0 & 16.1 - 15.5 & 3765 - 3686 & 3512 - 3171\\
 Miras - MS & 19.8 - 6.5 & 15.4 - 11.1 & 2795 - 826 & 1934 - 518 & 20.6 - 16.2 & 11.8 - 10.7 & 3313 - 2853 & 2364 - 1597\\
\hline 
\end{tabular}
\end{center}
\end{table*}

\begin{table*}
\caption[]{3.6m - astrophysical targets, J band. GS: Good seeing, MS: Median seeing. Strehls are in per cent.}
\label{adonis_astro_J}
\begin{center}
\begin{tabular}{|l||c|c|c|c||c|c|c|c|}
\hline 
 Object  &  $<S_{NGS}>$  & $\sigma_{S_{NGS}}$ & $N_{S>0.1}$ & $N_{S>0.2}$ & $<S_{LGS}>$ & $\sigma_{S_{LGS}}$ & $N_{S>0.1}$ & $N_{S>0.2}$ \\
\hline 
 Quasar - GS & 0.5 & 0.2 & 0 & 0 & 2.6 & 2.0 & 60 & 0\\
 Quasar - MS & 0.3 & 0.1 & 0 & 0 & 0.6 & 0.4 & 0 & 0 \\

 AGN - GS & 1.1 & 2.7 & 74 & 5 & 4.2 & 4.3 & 359 & 0\\
 AGN - MS & 0.4 & 0.7 & 1 & 0 & 1.1 & 1.2 & 0 & 0\\

 PMS - GS & 3.0 - 0.9 & 4.5 - 1.7 & 91 - 9 & 8 - 0 & 8.0 - 4.5 & 3.9 - 3.2 & 309 - 73 & 0 - 0\\
 PMS - MS & 1.1 - 0.4 & 2.1 - 0.5 & 12 - 0 & 0 - 0 & 2.0 - 1.0 & 1.2 - 0.8 & 0 - 0 & 0 - 0 \\

 SR - GS & 5.5 - 3.7 & 6.5 - 5.8 & 471 - 302 & 101 - 55 & 6.4 - 6.0 & 4.9 - 4.6 & 570 - 454 & 0 - 0\\
 SR - MS & 2.3 - 1.5 & 3.4 - 2.8 & 116 - 74 & 0 - 0 & 1.6 - 1.4 & 1.5 - 1.4 & 0 - 0 & 0 - 0 \\

 Miras - GS & 5.6 - 2.0 & 6.2 - 3.8 & 930 - 267 & 166 - 22 & 6.5 - 5.0 & 5.0 - 4.3 & 1239 - 647 & 0 - 0\\
 Miras - MS & 2.2 - 0.7 & 3.2 - 1.6 & 189 - 22 & 0 - 0 & 1.6 - 1.2 & 1.6 - 1.2 & 0 - 0 & 0 - 0\\ 
\hline 
\end{tabular}
\end{center}
\end{table*}

\begin{table*}
\caption[]{8m - astrophysical targets, K band. GS: Good seeing, MS: Median seeing. Strehls are in per cent.}
\label{naos_astro}
\begin{center}
\begin{tabular}{|l||c|c|c|c||c|c|c|c|}
\hline 
  Object  &  $<S_{NGS}>$  & $\sigma_{S_{NGS}}$ & $N_{S>0.1} $ & $N_{S>0.2} $ & $<S_{LGS}>$ & $\sigma_{S_{LGS}}$ & $N_{S>0.1}$ & $N_{S>0.2}$ \\
\hline 
 Quasar - GS & 4.4  & 7.2 & 819 & 357 & 37.2 & 17.0 & 7651 & 6803\\
 Quasar - MS & 1.2  & 2.0 & 63 & 21 & 16.7 & 10.6 & 5953 & 2893\\

 AGN  - GS & 15.4 & 23.1 & 895 & 698 & 43.0 & 23.3 & 2329 & 2077\\
 AGN - MS & 6.9 & 15.2 & 426 & 315 & 22.6 & 17.0 & 1850 & 1393\\

 PMS - GS & 45.2 - 19.8 & 27.1 - 23.1 & 771 - 439 & 710 - 317 & 63.0 - 50.0 & 14.0 - 19.2 & 918 - 868 & 892 - 822\\
 PMS - MS & 25.4 - 7.6 & 23.7 - 14.2 & 568 - 180 & 443 - 127 & 38.9 - 26.5 & 12.4 - 14.0 & 870 - 777 & 842 - 624\\

 SR - GS & 61.4 - 49.5 & 23.0 - 26.6 & 2020 - 1896 & 1961 - 1768 & 57.8 - 56.9 & 18.3 - 18.0 & 2050 - 2055 & 1984 - 1990\\
 SR - MS & 42.2 - 29.4 & 23.9 - 25.5 & 1835 - 1426 & 1663 - 1159 & 33.8 - 32.6 & 15.5 - 15.1 & 1877 - 1884 & 1721 - 1696 \\

 Miras - GS & 63.5 - 35.4 & 21.3 - 27.2 & 4084 - 3189 & 3961 - 2660 & 57.9 - 54.1 & 18.1 - 18.4 & 4096 - 4041 & 3960 - 3906\\
 Miras - MS & 44.2 - 17.4 & 22.4 - 22.0 & 3772 - 1811 & 3530 - 1342 & 33.8 - 29.7 & 15.4 - 14.9 & 3748 - 3679 & 3450 - 3148\\
\hline 
\end{tabular}
\end{center}
\end{table*}

\begin{table*}
\caption[]{8m - astrophysical targets, J band. GS: Good seeing, MS: Median seeing. Strehls are in per-cent.}
\label{naos_astro_J}
\begin{center}
\begin{tabular}{|l||c|c|c|c||c|c|c|c|}
\hline 
  Object  &  $<S_{NGS}>$  & $\sigma_{S_{NGS}}$ & $N_{S>0.1}$ & $N_{S>0.2}$ & $<S_{LGS}>$ & $\sigma_{S_{LGS}}$ & $N_{S>0.1} $ & $N_{S>0.2} $ \\
\hline 
 Quasar - GS & 0.7 & 1.5 & 43 & 12 & 11.0 & 7.6 & 4214 & 1119\\
 Quasar - MS & 0.2 & 0.2 & 1 & 0 & 2.1 & 1.9 & 40 & 0\\

 AGN  - GS & 4.9 & 12.2 & 350 & 241 & 15.4 & 12.5 & 1568 & 965\\
 AGN - MS & 1.6 & 5.3 & 148 & 68 & 3.6 & 3.9 & 287 & 0\\

 PMS - GS & 17.9 - 5.4 & 18.7 - 10.9 & 476 - 147 & 346 - 104 & 27.3 - 18.2 & 9.5 - 10.2 & 854 - 702 & 760 - 422\\
 PMS - MS & 6.5 - 1.3 & 10.6 - 3.9 & 203 - 35 & 124 - 14 & 7.1 - 3.9 & 3.5 - 2.99 & 196 - 46 & 0 - 0\\

 SR - GS & 30.4 - 21.1 & 20.2 - 20.6 & 1670 - 1213 & 1411 - 898 & 22.9 - 22.1 & 12.1 - 11.7 & 1736 - 1727 & 1406 - 1360 \\
 SR - MS &  13.7 - 8.7 & 13.9 - 12.9 & 1017 - 619 & 632 - 418 & 5.5 - 5.2 & 4.3 - 4.0 & 387 - 296 & 0 - 0\\

 Miras - GS & 31.7 - 12.4 & 19.2 - 0.17 & 3516 - 1497 & 2975 - 1045 & 23.0 - 19.8 & 12.2 - 11.5 & 3481 - 3231 & 2701 - 2271 \\
 Miras - MS & 14.1 - 4.2 & 13.3 - 9.0 & 2156 - 616 & 1294 - 371 & 5.6 - 4.4 & 4.4 - 3.8 & 868 - 428 & 0 - 0\\
\hline 
\end{tabular}
\end{center}
\end{table*}

Several  comments  can be made about  these  numbers.  The  first  is that the
improvement due to the laser star in K band is obvious: for faint objects, the
Strehl ratio and the number  of observable object  are notably increased.  The
gain of the  LGS is maximum for faint  objects (like quasars,  mean magnitude:
18.2, AGN: 16.7), for which the gain in  Strehl can be as  high as a factor of
10 (8m case).  Bright objects (SRs, at maximum (mean magnitude of 12.3), Miras
at maximum,  (mean magnitude: 12.6))   can be used  as  reference for  the NGS
system, which gives high Strehls (no cone  effect), and therefore benefit less
from the  laser star.  The  number of objects   is also increased considerably
with the LGS: for the 8m, the number of quasars goes from hundreds (21-357) to
thousands (2893-6803).  Another important  factor  is the influence  of seeing
conditions.  The fainter the objects, the more it  affects AO performance: for
the 3.6m, the number of quasars observable with the  LGS can range from 287 to
3809.  This is due to the  sensitivity of the LGS  to high altitude turbulence
and because for  these objects, the limiting  magnitude of the system is close
to the  magnitude of the objects  (see Fig.~\ref{fig:ao_lgs}).  On NGS AO, the
differences are  not as important (which confirms  the  dominant effect of the
cone effect).  This emphasizes again the need to be  able to estimate the high
altitude turbulence component.

Like in the statistical  approach, the results in J  are more contrasted.  One
can get a tremendous improvement on faint objects compared to  NGS (e.g.  from
12 to 1119 quasars on the 8m).  In almost all cases, when  the seeing is good,
the LGS improves  the situation, although  sometimes marginally (like  for the
SRs).    If the seeing   is  median, faint  objects   (like quasars, AGN)  are
difficult to observe with good performances (with $S>$0.2, no quasars, only 68
AGN for 8m NGS, and none with the LGS).  It is not unusual that better results
are obtained with NGS for median seeing, which is  the same conclusion as with
statistical methods, although some exceptions are present (8m, quasars and AGN
with $S>$0.1).  It appears that most of the time,  the best Strehl is achieved
on-axis, even if a reference star is  found, and can be  seen by the change in
Strehl  when going from  bright to faint phase of  variable stars.  Indeed, If
external  references were used in  a majority of cases,  the brightness of the
central object would not matter. The reason for this behavior can be explained
by several arguments. First, when looking at  Fig~\ref{fig:ao_lgs}, we can see
that the LGS  is at maximum Strehl  down to magnitude  15-16 (8m case), so  an
external reference is not needed at these magnitudes.  A second reason is that
since  science  catalogues are  incomplete, bright  objects will  dominate the
catalogue, and they can be observed  on-axis.  A third reason  is that for two
nearly   equivalent  brightness  objects,  the on-axis   one  will give better
performances, because there is no anisoplanatism.

\begin{figure}
\centerline{\psfig{figure=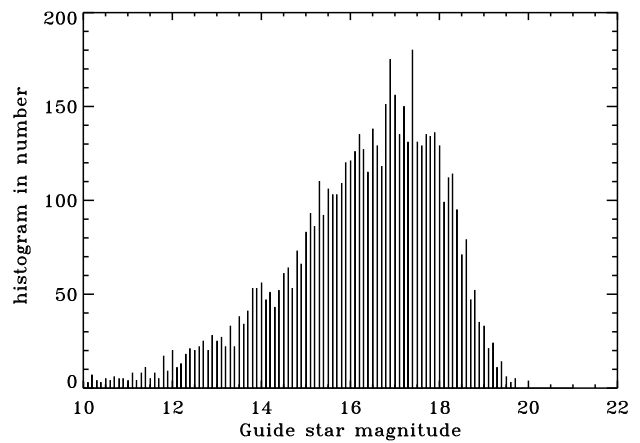,width=9.0cm}}
\caption{Histogram of the magnitude of the stars found in the USNOC, yielding the best Strehl when used as a reference}
\label{fig:usno_mag}
\end{figure}
A second point is the incompleteness of the USNOC.  The number of bright stars
will be over-estimated compared to  fainter ones, and therefore the references
will be  statistically further away  from the  science  object, yielding to an
under-estimation of Strehl. It also means that the numbers we have derived are
lower  limits, since  all the  stars  suitable as  references are   not in the
catalogue.  On  Fig.~\ref{fig:usno_mag},  we plotted  the  magnitudes  of  the
reference stars giving the best Strehl ratio.  It can be seen that stars close
to magnitude   20  were found in  this   catalogue, which   is a  considerable
improvement to the  Hubble Space Telescope  Guide Star  Catalogue (complete to
about magnitude   14.5,  ~\cite{Jenkner1990}) which  was  used  in  an earlier
version of this study.  This  histogram is a  convolution of two effects:  the
magnitude dependance  of the AO  system  (faint sources  are not  selected  as
references), and  the magnitude distribution  of the  catalogue.  The peak  of
Fig.~\ref{fig:usno_mag} is near magnitude 17, whereas the peak distribution of
star for the USNOC is near 19.  Therefore, this study  should suffer less from
incompleteness than when using the GSC.

The LGS can significantly increase the number  of objects that can be observed
in each class, so that statistical studies on these  objects can be performed.
The performances that   can be expected on individual   objects are also  very
promising.  On each  class of objects, one  can  find targets  that allow high
performance observations.  Indeed,  there are objects  with high Strehl ratios
and that could not be observed with the AO system without the LGS.  In J band,
the LGS improves the performances of the system, but only when the atmospheric
conditions are good.  The  number of objects  that  can be observed  with good
performances  is drastically reduced,   but in good seeing,  there  is still a
large sample of objects  (e.g.  1119 quasars,  965 AGN  with $S>0.2$)  to make
statistical studies.  With NGS,  this is not  as obvious (only 8  quasars, 128
AGN for the same  conditions).   However, even if  the  Strehl is  lower,  and
therefore the  gain in signal  to  noise ratio  obtained by the  correction is
smaller, the diffraction limit in J is also smaller,  so one can have a better
resolution.  Thus the LGS can be used in J band, under good seeing conditions,
to improve the performances of the AO system.

\section{LGS in the visible}
\begin{figure}
\centerline{\psfig{figure=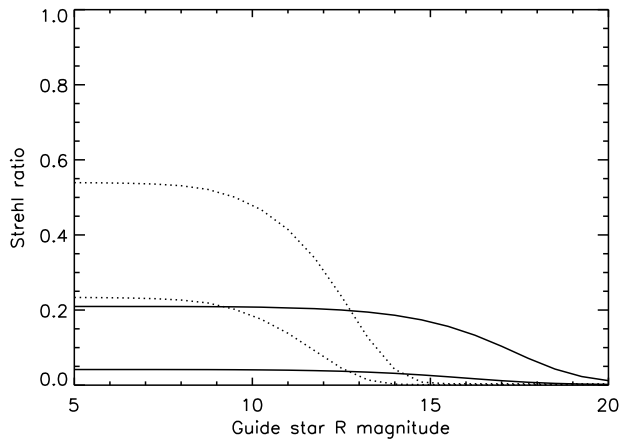,width=9.0cm}}
\caption{Strehl NGS (dots), LGS (solid), for excellent (NGS, LGS), and good seeing conditions, at $H_{\alpha}$.}
\label{fig:strehl_visible}
\end{figure}

To see  the  possibilities of  the   laser star in  the   visible part of  the
spectrum, we chose a excellent atmospheric model.  This model uses a seeing of
0.3 arcsec,  which  can be obtained approximately   10 per cent of   the time,
turbulence heights and wind profile being the same as in the good seeing model
previously used.  The  isoplanatic angle was chosen to  be larger  than in the
good model, 6.0  arcsec  (at  0.5  $\mu$m).  This was  done  by  modifying the
strength of  the layers.  Fig.~\ref{fig:strehl_visible}  represents the Strehl
with and   without a LGS, on   the  8m system,  as a   function of the on-axis
reference star R magnitude.  The solid curves are for the LGS AO, in excellent
and good seeing conditions.  The  dotted curves represent NGS  AO for the same
seeing   conditions.   It  can clearly  be  seen  that   when the  atmospheric
conditions are  excellent, but  not unrealistic, a   reasonable Strehl can  be
achieved with the laser star in the red.   At $\R{H}_\alpha$ (0.656 $\mu$m), a
Strehl of 0.2 can be achieved with the LGS down to magnitude 13-15, which is a
significant improvement.  The dotted  curve represents the performances of NGS
AO, in the same conditions.  Because there is no cone effect, the Strehl ratio
can be  as  high as 0.55  at  $\R{H}_\alpha$, in exceptional seeing.   At that
wavelength, the  diffraction limit of an  8m telescope is 17 milli-arcseconds,
which can be achieved with both  methods.  This is  nearly 4 times better than
the diffraction limit of the HST.  The gain  obtained with astronomical NGS AO
in the red has already been experimentally confirmed, on the CFHT AO bonnette,
PUEO ~\cite{Rigaut1997b}.

This result clearly shows that adaptive optics and LGS adaptive optics are not
limited to the infrared part of the spectrum as it  is often thought.

\section{Conclusions}

The  necessity   of   having  favorable atmospheric   conditions   to get good
performances with AO in general  and LGS AO in  particular underlines the need
for flexible scheduling. One has to be able to program high angular resolution
observations when the seeing is good.  A mean  of measuring, even crudely, the
$C_n^2$ profile on-line would be  of crucial importance  to be able to predict
if LGS observations are feasible, because of the height dependence of the cone
effect.  A SCIDAR  or a more simple device,  giving  measurements of the  high
altitude component of turbulence, would therefore be of prime importance.

We have shown  shown that good  performances  can be achieved  in the visible,
when atmospheric  conditions are good (1  night over 10) with adaptive optics.
It  is reminded that  astronomical adaptive optics is  not limited to the near
infra-red domain, as it is often thought in the astronomical community.

The  problem of  knowing what   Strehl  ratio yields acceptable  astrophysical
results should be studied in more detail. We chose to take  two cases, 0.1 and
0.2. However, it seems likely that with proper deconvolution methods, reliable
astrophysical results can   be obtained  with  lower Strehls.    The limit  is
probably not  sharp  and depends on the   class of object being observed,  but
deserves further investigation.  The   different behavior of the  point spread
function between NGS and LGS and its astrophysical implications should also be
studied.

This  study has shown the  imperative need, for  8m class telescopes to find a
solution   for  the cone  effect.   Indeed,   as shown   by both sky  coverage
approaches, a    good Strehl ratio    can be achieved on  a    large amount of
astrophysical objects in the  red, with the laser guide  star, only  with good
(or excellent) seeing, because of the cone effect.

The tilt determination problem is also important.  As shown by the statistical
approach, a full sky coverage cannot be achieved with the LGS.  Galactic poles
cannot be  fully exploited at  high angular resolution,  because of the lack of
nearby tilt-reference stars. To  achieve full  coverage, the only solution
is to  get  tilt  information from  the laser  star  (either the polychromatic
method or a combination of the laser-strip observing methods).

However, we  have  shown in  this  study, by  two different methods,  that the
simplest Laser Guide Star,  without correction of the cone  effect and using a
natural star for  tilt determination, can bring  a significant  improvement in
sky  coverage in   the K band,  both on   3.6m and  8m  telescopes.   We  made
approximations to correct for  the V-R values for  the observed objects, since
the wavefront  sensing    is done in   the  red.   For  an atmospheric   model
representing  the best 20 per  cent of the  time,  the sky  coverage is nearly
complete in the galactic plane for  Strehls of 0.3  (for the 3.6m) or 0.5 (for
the 8m).   At average galactic  latitudes, the coverage drops  to 25  per cent
(3.6m) and  30 per cent  (8m).  Near the  pole, the figures become small (0.08
and 0.7), because   tilt-reference stars are not  found  close to the  science
object.  We have also  shown that the sky  coverage and the improvement due to
the LGS  are very  sensitive to seeing   conditions.   For atmospheric  models
representing median values, the sky coverage values drop to 10 per cent (3.6m)
and 8  per  cent (8m) for  0.3  and 0.5 Strehl.  The  improvement  in terms of
observable objects is  also important.  The  number of observable quasars with
Strehl greater than 0.2 is increased from 23 to 3809 for the 3.6m and from 357
to 6803 on the 8m  (good seeing).  For  stellar  objects, whose population  is
dominated, in  the catalogues, by bright objects  do not benefit  as much from
the  LGS, since for example,  the semi-regular pulsating  variables, for which
the average  magnitude is as bright  as 12.3 at  maximum, are  better observed
with  the Natural guide star.   However, in a vast  majority of cases, the LGS
gives much better performances  than NGS in the K  band, the biggest increases
being  for the faintest  objects.  In J  band,  the improvement is smaller and
very  much seeing  dependent.    In good seeing,  the LGS   still improves the
performances of the system (e.g.  it brings 60 quasars in the 3.6m case, where
none were  accessible with NGS; for   the 8m, the  number  of AGN with Strehls
greater than 0.2 goes from 241 to 965).  However,  in median seeing, it can be
preferable to use  the NGS system, because  the cone effect reduces very  much
the performances of  the system (for example,  124 pre-main sequence stars can
be observed at 0.2 Strehl with the 8m-NGS, and none with the LGS).

\section*{ACKNOWLEDGEMENTS}

The authors would like  to thank A.  Robin,  from Observatoire de  Besan\c con
for providing the  stellar population models, F.  Rigaut  for providing us the
AO simulation code and for fruitful discussions, D.  Bonaccini for his help on
AO simulations, M.  Sarazin for  insightful discussions on atmospheric models,
and Eric  Thi\'ebaut for his  help.  This research  has made use of the SIMBAD
database, operated   at   CDS, Strasbourg, France,   and   of  the USNO-A-V1.0
catalogue. This work has benefitted   from discussions during meetings of  the
Laser Guide Star TMR network of the European Union, contract \#ERBFMRXCT960094.

\bsp
\label{lastpage}
\end{document}